\documentclass[lettersize,journal]{IEEEtran}
\usepackage{amsmath,amsfonts}
\usepackage{algorithmic}
\usepackage{algorithm}
\usepackage{array}
\usepackage{textcomp}
\usepackage{stfloats}
\usepackage{url}
\usepackage{verbatim}
\usepackage{graphicx}
\usepackage{cite}
\usepackage{amssymb}
\usepackage{setspace}
\usepackage{soul}
\usepackage{color}
\usepackage{subfigure}
\usepackage{pifont}
\usepackage[explicit]{titlesec}

\soulregister\cite7
\setstcolor{yellow}

\begin{document}

\title{Joint Beamforming for RIS-Assisted Integrated Sensing and Communication Systems}

\author{Yongqing Xu, \emph{Graduate Student Member, IEEE}, Yong Li, \emph{Member, IEEE}, \\J. Andrew Zhang, \emph{Senior Member, IEEE}, Marco Di Renzo, \emph{Fellow, IEEE}, and Tony Q. S. Quek, \emph{Fellow, IEEE}\\
	
        % <-this % stops a space
% \thanks{\emph{(Corresponding author: Yong Li)}}% <-this % stops a space
\thanks{The work of Y. Xu and Y. Li was supported by the National Research and Development Program ``Distributed Large Dimensional Wireless Cooperative Transmission Technology Research and System Verification'' under Grant 2022YFB2902400. The work of M. Di Renzo was supported in part by the European Commission through the H2020 ARIADNE project under grant agreement number 871464 and through the H2020 RISE-6G project under grant agreement number 101017011, and by the Agence Nationale de la Recherche (France 2030, ANR PEPR Future Networks, grant NF-YACARI, 22-PEFT-0005). The work of Tony Q. S. Quek was supported by the National Research Foundation, Singapore and Infocomm Media Development Authority under its Future Communications Research \& Development Programme. (Corresponding Author: Yong Li and Tony Q. S. Quek.)
	
Y. Xu and Y. Li are with the Key Laboratory of Universal Wireless Communications, Beijing University of Posts and Telecommunications, Beijing 100876, China. Email: \{xuyongqing; liyong\}@bupt.edu.cn. 

J. Andrew Zhang is with the Global Big Data Technologies Centre (GBDTC) and the School of Electrical and Data Engineering, The University of Technology Sydney (UTS), Sydney, NSW 2007, Australia. Email: Andrew.Zhang@uts.edu.au.

M. Di Renzo is with Universit\'e Paris-Saclay, CNRS, CentraleSup\'elec, Laboratoire des Signaux et Syst\`emes, 3 Rue Joliot-Curie, 91192 Gif-sur-Yvette, France. (marco.di-renzo@universite-paris-saclay.fr).

T. Q. S. Quek is with the Singapore University of Technology and Design, Singapore 487372, and also with the Yonsei Frontier Lab, Yonsei University, South Korea (e-mail: tonyquek@sutd.edu.sg).
}}%

%{Shell \MakeLowercase{\textit{et al.}}: A Sample Article Using IEEEtran.cls for IEEE Journals}%

%\IEEEpubid{0000--0000/00\$00.00~\copyright~2021 IEEE}%

\maketitle

\begin{abstract}
Integrated sensing and communications (ISAC) is an emerging technique for the next generation of communication systems. However, due to multiple performance metrics used for communication and sensing, the limited number of degrees-of-freedom (DoF) in optimizing ISAC systems poses a challenge. Reconfigurable intelligent surfaces (RISs) can introduce new DoF for beamforming in ISAC systems, thereby enhancing the performance of communication and sensing simultaneously. In this paper, we propose two optimization techniques for beamforming in RIS-assisted ISAC systems. The first technique is an alternating optimization (AO) algorithm based on the semidefinite relaxation (SDR) method and a one-dimension iterative (ODI) algorithm, which can maximize the radar mutual information (MI) while imposing constraints on the communication rates. The second technique is an AO algorithm based on the Riemannian gradient (RG) method, which can maximize the weighted ISAC performance metrics. Simulation results verify the effectiveness of the proposed schemes. The AO-SDR-ODI method is shown to achieve better communication and sensing performance, than the AO-RG method, at a higher complexity. It is also shown that the mean-squared-error (MSE) of the estimates of the sensing parameters decreases as the radar MI increases.
\end{abstract}


\section{Introduction}
\IEEEPARstart{E}{merging} applications in future networks, such as vehicle-to-everything and smart homes, demand high-quality communication and sensing capabilities \cite{ref1, renew4}. These requirements can be realized through a unified platform that combines communication and sensing functionalities, as they share a similar hardware architecture and signal processing algorithms. Motivated by these considerations, research on integrated sensing and communication (ISAC) is currently underway to empower next-generation communication systems with sensing capabilities, enabling dynamic spectrum sharing and harmonious co-existence between radar and communication systems \cite{ref3}.

Beamforming is a promising approach for realizing ISAC \cite{ref3,renew1}. In \cite{ref4}, the authors designed radar beamforming to project radar signals onto the null space of the interference channel, thereby enabling radar/communication coexistence. In \cite{ref5}, a robust multiple-input multiple-output (MIMO) beamforming scheme was proposed to realize vehicular communication and radar coexistence under imperfect channel state information (CSI). Additionally, the authors of \cite{ref6} proposed to use efficient manifold algorithms to solve the ISAC beamforming problem. The beamforming schemes in \cite{ref7,ref8} are based on orthogonal frequency-division multiplexing (OFDM). The authors of \cite{ref7} designed the transmit/receive radar beam patterns and the receive communication beam pattern to maximize the Kullback-Leibler divergence (KLD) while satisfying the signal-to-noise-plus-interference constraints of communication users. In \cite{ref8}, the authors proposed a novel adaptive OFDM waveform to maximize the radar mutual information (MI) and the communication rate. The radar MI leads to an optimum radar estimation waveform \cite{add1,add4}. Furthermore, the authors of \cite{ref9} designed a frequency-hopping waveform and developed accurate methods for estimating the timing offset and the channel based on the designed waveform. However, jointly designing communication and sensing performance metrics limits the degrees-of-freedom (DoF) of ISAC beamforming.

Reconfigurable intelligent surface (RIS) technology has garnered significant attention recently. An RIS is a planar surface composed of numerous low-cost and nearly passive reconfigurable elements that dynamically induce appropriate amplitude and phase shifts to the incident signals \cite{ref10, ref11, renew5}. By leveraging the RIS technology, the DoF for ISAC beamforming can be increased, and the signal coverage of RIS-assisted ISAC systems can be enhanced. Many papers on RISs for communication systems, covering joint beamforming \cite{ref12,ref14} and robust transmission design \cite{ref15,renew3}, have been published. However, RIS optimization requires knowledge of the channel state information (CSI) for both the base station (BS)-RIS and user equipment (UE)-RIS links \cite{ref16}. To obtain CSI, various methods have been proposed, including the discrete Fourier transform (DFT)-based method \cite{ref17}, compressed sensing (CS)-based channel estimation schemes \cite{ref19}, subspace-based estimation methods \cite{ref21}, and channel estimation algorithms based on machine learning \cite{ref24}. Furthermore, the deployment of RISs in radar systems has also been investigated. Specifically, the sensing performance in environments with rich scattering can be improved through RIS beamforming \cite{ref25,ref26,renew2}. For example, the authors of \cite{ref27} proposed a new self-sensing RIS architecture to improve the accuracy of angle estimation. RISs can also be deployed in scenarios where the infrastructure is insufficient for estimating the locations of UEs and scatterers. For example, with the use of RIS, theoretical bounds for localization were investigated in \cite{ref28}; signal strength based localization schemes were studied in \cite{ref29}; time-of-arrival (TOA) and angle of departure (AOD) based localization was investigated in \cite{ref30,ref31}.

Research on deploying RISs in ISAC systems is still in its infancy. To enable the co-existence of radar and communication functions, some researchers proposed to use RISs to minimize interference \cite{ref32}. In contrast, others suggested deploying two RISs near the transmitter and receiver to enhance the communication signals and suppress the mutual interference \cite{ref33}. Moreover, various studies have explored the design of ISAC signals, including increasing or maintaining the data rate \cite{ref30}, achieving a desired localization probability of error \cite{ref34}, enhancing the detection performance of ISAC systems in crowded areas while satisfying the communication signal-to-noise ratio (SNR) constraints \cite{ref35}, and minimizing multi-user interference \cite{ref36}. Some researchers have also explored the performance of ISAC systems in the terahertz (THz) band \cite{ref37}. In addition to passive RIS architectures, the authors of \cite{refHRIS1} proposed a novel hybrid reconfigurable intelligent surface (HRIS) architecture, which enables to reflect the impinging signal in a controllable manner, while simultaneously sense a portion of it. The authors of \cite{refHRIS2} further analyzed the advantages of applying HRISs in ISAC systems. However, the scenarios considered in previous studies are relatively simple, and no research has been conducted on designing a joint waveform based on the radar MI. MI can be used as a metric to measure both the radar and communication performance. The MI for communication systems quantifies the amount of information about the transmit signals when the channel is known. The MI for sensing systems, i.e., the radar MI, quantifies the amount of information about the channel when the transmit signals are known \cite{ref3}. By maximizing the MI between target response and target echoes, we can obtain the best estimation and classification capabilities \cite{add4}.

\subsection{The Contributions of This Work}
In this paper, we aim to investigate joint beamforming for RIS-assisted ISAC systems, considering the direct and multiple reflected signal links. We first establish signal models for the radar MI and the weighted user rate. Subsequently, we formulate two constrained optimization problems and propose two algorithms to solve them. Our proposed joint beamforming methods guarantee good performance for both sensing parameter estimation and communication rate simultaneously. In particular, even with multiple constraints on the communication users, the proposed methods can still achieve a small mean square error (MSE) for sensing parameter estimation.

The main contributions of our work are as follows:
\begin{itemize}
	\item 
	We establish a more practical signal model for RIS-assisted ISAC systems, considering the direct and multiple reflected signal links, based on which the radar MI and the weighted user rate are derived.
	
	\item We formulate a joint beamforming problem for maximizing the radar MI under specified constraints for the weighted user rate, transmit power, and discrete phase shifts at the RIS. Then, an alternating optimization algorithm based on the semidefinite relaxation and a one-dimension iterative (AO-SDR-ODI) algorithm is proposed to solve the formulated constrained optimization problem. Specifically, the ODI algorithm can solve the quartic RIS beamforming problem efficiently. The algorithm is shown to provide accurate solutions. However, its efficiency degrades and its complexity increases quickly when the number of RIS elements and users becomes very large.
	
	\item To solve these limitations of the AO-SDR-ODI algorithm, we propose another alternating optimization algorithm based on the Riemannian gradient (AO-RG) method. The AO-RG algorithm transforms the joint beamforming of the constrained optimization problem into a weighted optimization problem, efficiently solving optimization problems for RIS-assisted ISAC beamforming systems with many variables, i.e., the number of RIS elements and the number of users are large.
	
	\item Simulation results are presented to verify the effectiveness of the proposed AO-SDR-ODI and AO-RG algorithms. Specifically, the MSE of the sensing parameters decreases as the radar MI increases, indicating that the proposed algorithms can achieve better sensing performance. Moreover, the computational complexity of the proposed algorithms is analyzed.
\end{itemize}
\subsection{Paper Organization}
The remainder of this paper is organized as follows. Section \uppercase\expandafter{\romannumeral2} introduces the system model and studies relevant performance metrics. Section \uppercase\expandafter{\romannumeral3} formulates the optimization models for maximizing the radar MI and introduces the AO-SDR-ODI algorithm for solving it. Section \uppercase\expandafter{\romannumeral4} formulates a weighted optimization problem for maximizing the integrated radar and communication performance, and introduces the AO-RG algorithm to tackle it. Section \uppercase\expandafter{\romannumeral5} studies the complexity and convergence of the proposed algorithms. Section \uppercase\expandafter{\romannumeral6} presents the simulation results. Section \uppercase\expandafter{\romannumeral7} concludes the paper.
\subsection{Notations}
Throughout this paper, matrices are denoted by bold uppercase letters, e.g., $\boldsymbol{S}$, vectors are denoted by bold lowercase letters, e.g., $\boldsymbol{s}$, and scalars are denoted by normal fonts, e.g., $s$. $[\cdot]^T$, $[\cdot]^H$, and $[\cdot]^*$ stand for the transpose, the complex conjugate transpose, and the complex conjugate operations. $\mathcal{CN}$ denotes the circularly-symmetric Gaussian random distribution. $\mathbb{C}^{\left(\cdot\right)}$ denotes the complex space. $\mathbb{E}\left[\cdot\right]$ denotes the statistical expectation. $I(\boldsymbol{x};\boldsymbol{y}|\boldsymbol{z})$ denotes the MI between $\boldsymbol{x}$ and $\boldsymbol{y}$ under the condition that the random vector $\boldsymbol{z}$ is given. $h(\cdot)$ and $h(\cdot|\cdot)$ represent the differential entropy and the conditional differential entropy, respectively. $p(\cdot)$ represents the probability density function. $\log(\cdot)$ is the natural logarithm. $\det(\cdot)$ and $\text{tr}(\cdot)$ represent the determinant and trace of a matrix, respectively. $\text{diag}(\boldsymbol{s})$ represents a matrix whose diagonal entries are composed of $\boldsymbol{s}$. $\text{diag}(\boldsymbol{S})$ represents a vector whose entries are composed of the diagonal entries of $\boldsymbol{S}$. $\Vert\cdot\Vert$ stands for the $l_2$ norm. $(s)^+$ stands for the positive part of $s$, i.e, $(s)^+=\text{max}(s,0)$. $\text{rank}(\cdot)$ stands for the rank of a matrix. $\text{Re}(\cdot)$ stands for the real part of the argument. $\circ$ stands for the Hadamard product. $\left\langle\cdot,\cdot\right\rangle$ stands for the Euclidean inner product.
\section{System Model}
\begin{figure}[!t]
	\centering
	\includegraphics[width=0.45\textwidth]{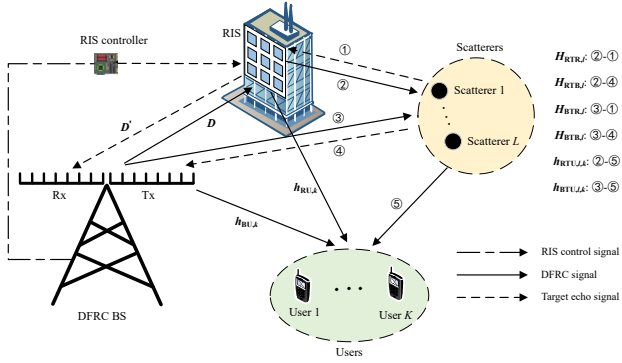}
	\caption{Considered RIS-assisted ISAC system. There are four signal paths between the BS and each scatterer, including the BS-RIS-scatterer-RIS-BS path, the BS-scatterer-BS path, the BS-scatterer-RIS-BS path, and the BS-RIS-scatterer-BS path. There are five signal paths between the BS and each UE, including the BS-UE path, the BS-RIS-UE path, the BS-scatterers-UE path, the BS-RIS-scatterers-UE path, and the BS-scatterers-RIS-UE path. Moreover, the notation \ding{173}-\ding{172} represents the two-way target scattered channel, consisting of the links \ding{173} and \ding{172}.}
	\label{fig_1}
\end{figure}

The considered system model is depicted in Fig. \ref{fig_1}, where a dual-functional radar and communication (DFRC) BS with $ N_t $ transmit antennas and $ N_r $ receive antennas communicates with $ K $ single-antenna users while sensing some scattering targets, with the aid of an $ M $-element RIS. The transmit and receive antennas are assumed uniform linear arrays (ULAs). The inter-distance of the transmit antennas, receive antennas, and RIS elements are equal to half of the signal wavelength. The transmit and receive antennas at the BS are widely separated to suppress the leakage signals between them. Additionally, $ L $ scatterers are uniformly distributed in the proximity of one scatterer center\footnote{We neglect the signals scattered twice by the scatterers, as their strength is much weaker compared to other signals. Moreover, it is noteworthy that, in practical scenarios, the spatial distribution of the scatterers may be unknown. However, we assume that beam tracking \cite{ref38} has been performed to detect the scatterers, and further observations are required to estimate other parameters such as the distance, velocity, and angle.}. The sets of users, RIS elements, and scatterers are denoted by $ \mathcal{K} $, $ \mathcal{M} $, and $ \mathcal{L} $, respectively. The phases of the RIS elements are controlled by an RIS controller, which communicates with the DFRC BS for cooperative transmission \cite{ref12}. 
\subsection{Signal Model}
The transmit beamforming matrix of the DFRC BS is $ \boldsymbol{S}=\left[  \boldsymbol{s}_1, \boldsymbol{s}_2, \cdots, \boldsymbol{s}_K\right] \in \mathbb{C}^{N_t \times K} $, where the transmit beamforming vector of the $ k $-th user is denoted by $ \boldsymbol{s}_k=\left[ s_{1,k}, s_{2,k}, \cdots, s_{N_t,k}\right]^T \in \mathbb{C}^{N_t \times 1} $. The transmit signal matrix is $ \boldsymbol{X}=\left[  \boldsymbol{x}_1, \boldsymbol{x}_2, \cdots, \boldsymbol{x}_K\right]^T \in \mathbb{C}^{K \times N} $, where the transmit signal of the $ k $-th user is denoted by $ \boldsymbol{x}_k=\left[ x_{k,1}, x_{k,2}, \cdots, x_{k,N}\right]^T\in \mathbb{C}^{N \times 1} $ with $N$ denoting the number of signal samples. The signal vectors in $\boldsymbol{X}$ are assumed to be statistically orthogonal to each other, so we have $\mathbb{E}\left[\boldsymbol{X}\boldsymbol{X}^H\right]=\boldsymbol{I}_K$. Then, the received signals at the receive antennas of the DFRC BS can be expressed as\footnote{We assume that all scatterers belong to the same extended target. As our system is narrowband, we consider only the angular spread of the extended target and do not account for the time-delay spread, i.e., the time-delays of all the scatterers are located in the same range bin \cite{reftimedelay}.}$^,$\footnote{The time-delay differences between different signal paths are much smaller than the signal sampling time in the considered system, and hence will not cause notable radar MI degradation. In addition, the time delays can typically be estimated and compensated if desired, as stated in \cite{ref27,zhang2022joint}. Therefore, we ignore the performance degradation that may be caused by differences in the time-delays of the signal paths.}
\begin{equation}
	\label{eq1}
	\begin{aligned}
	\boldsymbol{Y}_{\text{s}} &= \sum_{l=1}^{L}\Big[\left(\boldsymbol{D\Theta}\boldsymbol{H}_{\text{RTR},l}\boldsymbol{\Theta}\boldsymbol{D}^{\prime} \right)^H+\boldsymbol{H}_{\text{BTB},l}+\left(\boldsymbol{H}_{\text{BTR},l}\boldsymbol{\Theta D}^{\prime}\right)^H\\
	&\quad\quad+\left(\boldsymbol{D\Theta}\boldsymbol{H}_{\text{RTB},l}\right)^H\Big]\boldsymbol{S}\boldsymbol{X} + \boldsymbol{W},
	\end{aligned}
\end{equation}
where $ \boldsymbol{Y}_{\text{s}} = \left[ \boldsymbol{y}_{\text{s},1}, \cdots, \boldsymbol{y}_{\text{s},N_r} \right]^T \in \mathbb{C}^{N_r \times N} $ is the matrix of the received signals and $\boldsymbol{y}_{\text{s},n}\in \mathbb{C}^{N \times 1}$ is the received signal samples at the $n$-th antenna, and $ \boldsymbol{D} \in \mathbb{C}^{N_t \times M} $ is the channel matrix between the transmit antennas of the DFRC BS and the RIS, which can be written as 
\begin{equation}
	\label{eq2}
	\boldsymbol{D} = \left[ \boldsymbol{d}_1^H,\boldsymbol{d}_2^H,\cdots,\boldsymbol{d}_k^H,\cdots,\boldsymbol{d}_{N_t}^H  \right],
\end{equation}
where $\boldsymbol{d}_k \in \mathbb{C}^{M \times 1} $ is the channel vector between the $ k $-th transmit antenna and the RIS elements, and $\boldsymbol{D} = \beta_{\text{BR}}\boldsymbol{G}_{\text{BR}}$ with $\beta_{\text{BR}} = \sqrt{\frac{\lambda^2}{16\pi^2(d_{\text{BI}})^2}}$ denoting the large-scale path loss and $\boldsymbol{G}_{\text{BR}}$ denoting the small-scale channel matrix, where $d_{\text{BR}}$ is the distance between the transmit antennas array at the BS and the RIS. Moreover, $ \boldsymbol{D}^{\prime} \in \mathbb{C}^{M \times N_r} $ is the channel matrix between the receive antennas of the DFRC BS and the RIS elements, which can be written as
\begin{equation}
	\label{eq3}
	\boldsymbol{D}^{\prime} = \left[ \left(\boldsymbol{d}_1^{\prime}\right)^H,\left(\boldsymbol{d}_2^{\prime}\right)^H,\cdots,\left(\boldsymbol{d}_m^{\prime}\right)^H,\cdots,\left(\boldsymbol{d}_M^{\prime}\right)^H  \right],
\end{equation}
where $ \boldsymbol{d}_m^{\prime} \in \mathbb{C}^{N_r \times 1} $ is the channel vector between the $ m $-th RIS element and the receive antennas, and $\boldsymbol{D}^{\prime} = \beta_{\text{RB}}\boldsymbol{G}_{\text{RB}}$ with $\beta_{\text{RB}} = \sqrt{\frac{\lambda^2}{16\pi^2(d_{\text{RB}})^2}}$ denoting the large-scale path loss and $\boldsymbol{G}_{\text{RB}}$ denoting the small-scale channel matrix, where $d_{\text{RB}}$ is the distance between the RIS and the BS receive antennas. In addition, $ \boldsymbol{\Theta} $ denotes the beamforming matrix of the RIS, which is given by 
\begin{equation}
	\label{eq4}
	\boldsymbol{\Theta} = \text{diag}\left( \theta_1^H,\theta_2^H,\cdots, \theta_m^H, \cdots,\theta_M^H \right),
\end{equation}
where $ \theta_m $ is the phase shift of the $ m $-th RIS element. We consider both an ideal RIS reflection magnitude model, where each element fulfills the condition $\vert\theta_m\vert=1,\forall m\in\mathcal{M}$, and a practical RIS reflection phase shift model where each $\theta_m$ can only take $d$ finite values, equally spaced in $[0, 2\pi)$. The discrete set of phase shifts for each RIS element is given by 
\begin{equation} 
	\label{eqadd1}
	\mathcal{F} \triangleq \{\theta | \theta = e^{(j\frac{2\pi}{d}i + \frac{\pi}{d})}, 
	i = 0, ..., d-1\}. 
\end{equation} 
Moreover, $ \boldsymbol{H}_{\text{RTR},l} \in \mathbb{C}^{M \times M} $ is the backscattered matrix of the link between the $ l $-th scatterer and the RIS, which is assumed to have a circularly-symmetric complex Gaussian (CSCG) distribution according to \cite{ref39}, i.e., $ \boldsymbol{H}_{\text{RTR},l} \sim \mathcal{CN}\left( 0, \boldsymbol{R}_{\text{RTR},l}\right) $, and $\boldsymbol{H}_{\text{RTR},l} = \alpha_{\text{RTR},l}\boldsymbol{G}_{\text{RTR},l}$ with $\alpha_{\text{RTR},l} = \sqrt{\frac{\lambda^2\kappa}{64\pi^3(d_{\text{RT},l})^4}}$ denoting the large-scale path loss and $\boldsymbol{G}_{\text{RTR},l}$ denoting the small-scale channel matrix, where $\lambda$ is the wavelength, $\kappa$ is the radar cross-section (RCS) of the scatterers, and $d_{\text{RT},l}$ is the distance between the RIS and the $l$-th scatterer; $ \boldsymbol{H}_{\text{BTB},l} \in \mathbb{C}^{N_r \times N_t} $ is the backscattered matrix of the link between the $ l $-th scatterer and the BS, $ \boldsymbol{H}_{\text{BTB},l} \sim \mathcal{CN}\left( 0, \boldsymbol{R}_{\text{BTB},l}\right) $, and $\boldsymbol{H}_{\text{BTB},l} = \alpha_{\text{BTB},l}\boldsymbol{G}_{\text{BTB},l}$ with $\alpha_{\text{BTB},l} = \sqrt{\frac{\lambda^2\kappa}{64\pi^3(d_{\text{BT},l})^4}}$ denoting the large-scale path loss and $\boldsymbol{G}_{\text{BTB},l}$ denoting the small-scale channel matrix, where $d_{\text{BT},l}$ is the distance between the BS and the $l$-th scatterer; $ \boldsymbol{H}_{\text{BTR},l} \in \mathbb{C}^{N_t \times M} $ is the backscattered matrix of the link between the BS, the $ l $-th scatterer, and the RIS, $ \boldsymbol{H}_{\text{BTR},l} \sim \mathcal{CN}\left( 0, \boldsymbol{R}_{\text{BTR},l}\right) $, and $\boldsymbol{H}_{\text{BTR},l} = \alpha_{\text{BTR},l}\boldsymbol{G}_{\text{BTR},l}$ with $\alpha_{\text{BTR},l} = \sqrt{\frac{\lambda^2\kappa}{64\pi^3(d_{\text{BT},l})^2(d_{\text{RT},l})^2}}$ denoting the large-scale path loss and $\boldsymbol{G}_{\text{BTR},l}$ denoting the small-scale channel matrix; $\boldsymbol{H}_{\text{RTB},l}$ is the backscattered matrix of the link between the RIS, the $ l $-th scatterer, and the BS, and $ \boldsymbol{H}_{\text{RTB},l} = \boldsymbol{H}_{\text{BTR},l}^T$. In addition, $ \boldsymbol{W} $ is the additive white Gaussian noise (AWGN) at the receive antennas of the DFRC BS, with $ \boldsymbol{W} \sim \mathcal{CN}\left( 0, \boldsymbol{R}_\text{W}\right) $, and $\boldsymbol{R}_\text{W} \in \mathbb{C}^{N \times N}$ is a diagonal matrix.\\
\textbf{\emph{Remark 1:}} \emph{The received signals in (1) consist of five components, including the RIS-scatterers-RIS links, the BS-scatterers-RIS links, the BS-scatterers-RIS cascaded links, the RIS-scatterers-BS cascaded links, and the AWGN. It is reasonable to assume that there are line-of-sight (LOS) links between the BS and the RIS, as well as between the RIS and the scatterers, provided that the RIS is appropriately deployed.}\par
The received signals at the $ k $-th user can be formulated as
\begin{equation}
	\label{eq5}	
	\begin{aligned}
		&\boldsymbol{y}_{\text{c},k}=\Bigg[\boldsymbol{h}_{\text{BU},k} +  	\boldsymbol{D\Theta}\boldsymbol{h}_{\text{RU},k}\\
		&+\sum_{l=1}^{L}\big(\boldsymbol{h}_{\text{BTU},l,k}+\boldsymbol{D\Theta h}_{\text{RTU},l,k}+\boldsymbol{H_{\text{BTR},l}\Theta h_{\text{RU},k}}\big) \Bigg]^H\boldsymbol{s}_k\boldsymbol{x}_k^T\\
		&+ \sum_{i=1,i \neq k}^{K}\Bigg[ \boldsymbol{h}_{\text{BU},k} + \boldsymbol{D\Theta h}_{\text{RU},k}\\
		&+\sum_{l=1}^{L}\big(\boldsymbol{h}_{\text{BTU},l,k}+\boldsymbol{D\Theta h}_{\text{RTU},l,k}+\boldsymbol{H_{\text{BTR},l}\Theta h_{\text{RU},k}}\big)\Bigg]^H\boldsymbol{s}_i\boldsymbol{x}_i^T\\
		&+\boldsymbol{u}_k, \forall k \in \mathcal{K}, 
	\end{aligned}
\end{equation}
\noindent where $\boldsymbol{h}_{\text{BU},k} \in \mathbb{C}^{N_t \times 1}$ is the channel vector between the BS and the $k$-th user, and $\boldsymbol{h}_{\text{BU},k} = \beta_{\text{BU},k}\boldsymbol{g}_{\text{BU},k}$ with $\beta_{\text{BU},k} = \sqrt{\frac{\lambda^2}{16\pi^2(d_{\text{BU},k})^2}}$ denoting the large-scale path loss and $\boldsymbol{g}_{\text{BU},k}$ is assumed to follow the Rician fading channel model, i.e., $\boldsymbol{g}_{\text{BU},k} = \sqrt{\frac{\gamma_{\text{BU},k}}{1+\gamma_{\text{BU},k}}}\boldsymbol{g}_{\text{BU},k}^{\text{LOS}}+\sqrt{\frac{1}{1+\gamma_{\text{BU},k}}}\boldsymbol{g}_{\text{BU},k}^{\text{NLOS}}$, where $\boldsymbol{g}_{\text{BU},k}^{\text{LOS}}$ is the deterministic line-of-sight (LOS) component, $\boldsymbol{g}_{\text{BU},k}^{\text{NLOS}}$ is the random Rayleigh fading component, and $\gamma_{\text{BU},k}$ is the Rician factor; $\boldsymbol{h}_{\text{RU},k} \in \mathbb{C}^{M \times 1}$ is the channel vector between the RIS and the $k$-th user, and $\boldsymbol{h}_{\text{RU},k} = \beta_{\text{RU},k}\boldsymbol{g}_{\text{RU},k}$ with $\beta_{\text{RU},k} = \sqrt{\frac{\lambda^2}{16\pi^2(d_{\text{RU},k})^2}}$ denoting the large-scale path loss and $\boldsymbol{g}_{\text{RU},k} = \sqrt{\frac{\gamma_{\text{RU},k}}{1+\gamma_{\text{RU},k}}}\boldsymbol{g}_{\text{RU},k}^{\text{LOS}}+\sqrt{\frac{1}{1+\gamma_{\text{RU},k}}}\boldsymbol{g}_{\text{RU},k}^{\text{NLOS}}$. Moreover, $\boldsymbol{h}_{\text{BTU},l,k}\in \mathbb{C}^{N_t\times1}$ is the backscattered vector of the link between the BS, the $l$-th scatterer, and the $k$-th user, $\boldsymbol{h}_{\text{BTU},l,k}\sim\mathcal{CN}\left(0,\boldsymbol{R}_{\text{BTU},l,k}\right)$, and $\boldsymbol{h}_{\text{BTU},l,k}=\alpha_{\text{BTU},l,k}\boldsymbol{g}_{\text{BTU},l,k}$ with $\alpha_{\text{BTU},l,k}=\sqrt{\frac{\lambda^2\kappa}{64\pi^3(d_{\text{BT},l})^2(d_{\text{TU},l,k})^2}}$ and $\boldsymbol{g}_{\text{BTU},l,k}$ denoting the small-scale channel vector, where $d_{\text{TU},l,k}$ is the distance between the $l$-th scatterer and the $k$-th user; $\boldsymbol{h}_{\text{RTU},l,k}\in \mathbb{C}^{M\times1}$ is the backscattered vector of the link between the RIS, the $l$-th scatterer, and the $k$-th user, $\boldsymbol{h}_{\text{RTU},l,k}\sim\mathcal{CN}\left(0,\boldsymbol{R}_{\text{RTU},l,k}\right)$, and $\boldsymbol{h}_{\text{RTU},l,k}=\alpha_{\text{RTU},l,k}\boldsymbol{g}_{\text{RTU},l,k}$ with $\alpha_{\text{RTU},l,k}=\sqrt{\frac{\lambda^2\kappa}{64\pi^3(d_{\text{RT},l})^2(d_{\text{TU},l,k})^2}}$ and $\boldsymbol{g}_{\text{BTU},l,k}$ denoting the small-scale channel vector.\\
\textbf{\emph{Remark 2:}} \emph{The received signals of the $k$-th user consist of three parts: the useful signals given by the BS direct links and the BS-RIS-UE links, the interfering signals from other users on these links, and the AWGN. We assume that all the communication channels are perfectly estimated, and the sample covariance matrices for sensing are perfectly known as well\footnote{The unknown parameters are assumed to have been estimated during the searching and estimation stage (Stage 1) \cite{ref1}. Therefore, the sample covariance matrices and the communication channels are assumed to be known during the beamforming stage (Stage 2).}. However, obtaining accurate communication channel vectors, matrices, and sensing sample covariance matrices in practical scenarios is challenging. The analysis of these aspects is postponed to a future research work.}

\subsection{Radar Mutual Information}
Assume that the transmitted waveform $ \boldsymbol{SX} $ is known \cite{ref39}. The MI between the channel of the scatterers and the received echo can be expressed as
\begin{equation}
	\label{eq6}
\begin{aligned}
	I\left( \boldsymbol{Y}_{\text{s}};\boldsymbol{H}_{\text{s}} \vert \boldsymbol{SX} \right) &\triangleq h\left( \boldsymbol{Y}_{\text{s}} \vert \boldsymbol{SX} \right) - h\left( \boldsymbol{Y}_{\text{s}} \vert \boldsymbol{H}_{\text{s}}, \boldsymbol{SX} \right) \\
	& = h\left( \boldsymbol{Y}_{\text{s}} \vert \boldsymbol{SX} \right) - h\left( \boldsymbol{W} \right),
\end{aligned}
\end{equation}
where $\boldsymbol{H}_{\text{s}}$ is the channel of the scatterers. The differential entropy $ h\left( \boldsymbol{Y}_{\text{s}} \vert \boldsymbol{SX} \right) $ and $ h\left( \boldsymbol{W} \right) $ are derived in Appendix A. Accordingly, the radar MI $ I\left( \boldsymbol{Y}_{\text{s}};\boldsymbol{H}_{\text{s}} \vert \boldsymbol{SX} \right) $ can be expressed as
\begin{equation}
	\label{eq7}
	\begin{aligned}
	I\left( \boldsymbol{Y}_{\text{s}};\boldsymbol{H}_{\text{s}} \vert \boldsymbol{SX} \right) &= N_r\Bigg\{ \log\Bigg[\det\left(\sum_{l=1}^{L}\boldsymbol{\Lambda}_l + \boldsymbol{R}_W\right)\Bigg]\\
	&\quad-\log\left[\det\left(\boldsymbol{R}_W\right)\right] \Bigg\},
	\end{aligned}
\end{equation}
where
\begin{equation}
	\label{eq8}
	\begin{aligned}
	\boldsymbol{\Lambda}_l = \boldsymbol{X}^H\boldsymbol{S}^H\left(\hat{\boldsymbol{R}}_{\text{RTR},l}+\hat{\boldsymbol{R}}_{\text{BTB},l}+\hat{\boldsymbol{R}}_{\text{BTR},l}+\hat{\boldsymbol{R}}_{\text{RTB},l}\right)\boldsymbol{SX}&,\\
	\forall l \in \mathcal{L}&,
	\end{aligned}
\end{equation}
and
\begin{equation}
	\label{eq9}
	\begin{aligned}
		\left\{
		\begin{aligned}
		&\hat{\boldsymbol{R}}_{\text{RTR},l} = \boldsymbol{D\Theta H}_{\text{RTR},l}\boldsymbol{\Theta D}^{\prime}\left(\boldsymbol{D}^{\prime}\right)^H\boldsymbol{\Theta}^H\boldsymbol{H}_{\text{RTR},l}^H\boldsymbol{\Theta}^H\boldsymbol{D}^H, \\
		&\hat{\boldsymbol{R}}_{\text{BTB},l} = \boldsymbol{H}_{\text{BTB},l}^H\boldsymbol{H}_{\text{BTB},l},\\
		&\hat{\boldsymbol{R}}_{\text{BTR},l} = \boldsymbol{H}_{\text{BTR},l}\boldsymbol{\Theta D}^{\prime}\left(\boldsymbol{D}^{\prime}\right)^H\boldsymbol{\Theta}^H\boldsymbol{H}_{\text{BTR},l}^H,\\
		&\hat{\boldsymbol{R}}_{\text{RTB},l} = \boldsymbol{D\Theta H}_{\text{RTB},l}\boldsymbol{H}_{\text{RTB},l}^H\boldsymbol{\Theta}^H\boldsymbol{D}^H,
		\end{aligned}
		\right.
	\end{aligned}
\end{equation}
are sample covariance matrices.

\subsection{Weighted User Rate}
The average power of the effective signals for the $k$-th user can be formulated from (\ref{eq5}) as 
\begin{equation}
	\label{eq10}
	\begin{aligned}
		\xi_{k} =& \Bigg|\Bigg[\boldsymbol{h}_{\text{BU},k} +  	\boldsymbol{D\Theta}\boldsymbol{h}_{\text{RU},k}+\sum_{l=1}^{L}\big(\boldsymbol{h}_{\text{BTU},l,k}+\boldsymbol{D\Theta h}_{\text{RTU},l,k}\\
		&+\boldsymbol{H_{\text{BTR},l}\Theta h_{\text{RU},k}}\big) \Bigg]^H\boldsymbol{s}_k \Bigg|^2,
	\end{aligned}
\end{equation}
where the average signal power of the $k$-th user $\left\Vert\boldsymbol{x}_k\right\Vert^2$ is equal to one. The average interference plus noise power for the $ k $-th user can also be formulated from (\ref{eq5}) as 
\begin{equation}
	\label{eq11}
	\begin{aligned}
		\zeta_{k}=& \Bigg|\sum_{i=1,i \neq k}^{K}\Bigg[ \boldsymbol{h}_{\text{BU},k} + \boldsymbol{D\Theta h}_{\text{RU},k}+\sum_{l=1}^{L}\big(\boldsymbol{h}_{\text{BTU},l,k}\\
		&+\boldsymbol{D\Theta h}_{\text{RTU},l,k}+\boldsymbol{H_{\text{BTR},l}\Theta h_{\text{RU},k}}\big)\Bigg]^H\boldsymbol{s}_i\Bigg|^2 + \sigma_{u_{k}}^{2},
	\end{aligned} 
\end{equation}
where $\sigma_{u_{k}}^{2}$ is the variance of the AWGN $\boldsymbol{u}_k$. 
Then, the weighted communication rate for the $k$-th user \cite{ref40} can be written as
\begin{equation}
	\label{eq12}
	R_{k}=\alpha_{k} \log \left(1+\frac{\xi_{k}}{\zeta_{k}}\right), \forall k \in \mathcal{K},
\end{equation}
where $ \alpha_{k} $ is the weighting factor for the $k$-th user.
\section{Joint Beamforming for Radar Mutual Information Maximization}
This section aims to maximize the radar MI while considering the communication rate. It is worth noting that the radar MI and the communication rate depend on the transmit and RIS beamforming. Also, the radar MI maximization problem is challenging to solve due to the coupling between the transmit and RIS beamforming and the quartic expression of the radar MI. In order to tackle this problem, an AO-SDR-ODI algorithm is proposed. 
\subsection{Problem Formulation}
The beamforming optimization problem for maximizing the radar MI is formulated as
\begin{subequations}
	\label{eq13}
	\begin{align}
		\max_{\boldsymbol{S}, \boldsymbol{\Theta}} \quad & \log\left[\det\left(\sum_{l=1}^{L}\boldsymbol{\Lambda}_l + \boldsymbol{R}_W\right)\right] \label{eq13a} \tag{14a} \\
		\text{s.t.} \quad &  R_{k} \geq R_{0}, \forall k \in \mathcal{K}, \label{eq13b} \tag{14b} \\
		&\operatorname{tr}\left(\boldsymbol{S}\boldsymbol{S}^{H}\right) \leq P_{0}, \label{eq13c} \tag{14c} \\
		&\boldsymbol{\Theta}_{m,m}\in\mathcal{F}, \forall m \in \mathcal{M}. \label{eq13d} \tag{14d}
	\end{align}
\end{subequations}
In the formulated problem, the constant terms, i.e., $N_r$ and $ \log \left[\det\left(\boldsymbol{R}_W\right)\right]$, are omitted; (\ref{eq13b}) is the weighted user rate constraint, and $ R_0 $ is the minimum required user rate; (\ref{eq13c}) is the total transmit power constraint, and $ P_0 $ is the transmit power budget; (\ref{eq13d}) is the discrete phase shift constraint for the reflection coefficients at the RIS. Note that the constraints (\ref{eq13b}) and (\ref{eq13d}) are non-convex, and the transmit beamforming and the RIS beamforming are coupled in (\ref{eq13a}) and (\ref{eq13b}). Hence, the radar MI maximization problem is non-trivial to solve. Therefore, we propose an AO algorithm to tackle the problem in (\ref{eq13a}), which solves the transmit beamforming subproblem and the RIS beamforming subproblem iteratively.
\vspace{-0.8cm}
\subsection{Transmit Beamforming}
In this subsection, the RIS beamforming matrix is fixed. The transmit beamforming problem is formulated as a semidefinite programming (SDP) problem that can be solved via the SDR method. We first transform the radar MI into an equivalent form that can be written as\footnote{Assuming a sufficiently large number of signal samples, $\boldsymbol{XX}^H = \boldsymbol{I}_N$ holds, which is a reasonable assumption as hundreds or thousands of signal samples can be collected within one radar coherent processing duration.}
\begin{equation}
	\label{eq14}
	\begin{aligned}
		&\det\left(\boldsymbol{X}^H\boldsymbol{S}^H\boldsymbol{R}_{\text{s}}\boldsymbol{SX}+\sigma_{\text{s}}^2\boldsymbol{I}_N\right)\\
		&=\det\left[
		\begin{pmatrix}
			\boldsymbol{S}^H\boldsymbol{R}_{\text{s}}\boldsymbol{SX}\boldsymbol{X}^H,& \boldsymbol{0}_{K,N-K}\\
			\boldsymbol{0}_{N-K,K},& \boldsymbol{0}_{N-K,N-K}
		\end{pmatrix}
		+\sigma_{\text{s}}^2\boldsymbol{I}_N\right]\\
		&=\det\left[
		\begin{pmatrix}
			\boldsymbol{S}^H\boldsymbol{R}_{\text{s}}\boldsymbol{S},& \boldsymbol{0}_{K,N-K}\\
			\boldsymbol{0}_{N-K,K},& \boldsymbol{0}_{N-K,N-K}
		\end{pmatrix}
		+\sigma_{\text{s}}^2\boldsymbol{I}_N\right],
	\end{aligned}
\end{equation}
where $\boldsymbol{R}_{\text{s}} = \sum_{l=1}^{L}\left(\hat{\boldsymbol{R}}_{\text{RTR},l}+\hat{\boldsymbol{R}}_{\text{BTB},l}+\hat{\boldsymbol{R}}_{\text{BTR},l}+\hat{\boldsymbol{R}}_{\text{RTB},l}\right)$, and $\sigma_{\text{s}}^2$ is the variance of the AWGN $\boldsymbol{W}$. Equation (\ref{eq14}) can be derived using Sylvester's criterion. Furthermore, by utilizing the Hadamard inequality, an upper bound for (\ref{eq14}) is expressed as
\begin{equation}
	\label{eq15}
	\begin{aligned}
	&\det\left[
	\begin{pmatrix}
		\boldsymbol{S}^H\boldsymbol{R}_{\text{s}}\boldsymbol{S},& \boldsymbol{0}_{K,N-K}\\
		\boldsymbol{0}_{N-K,K},& \boldsymbol{0}_{N-K,N-K}
	\end{pmatrix}
	+\sigma_{\text{s}}^2\boldsymbol{I}_N\right] \\
	&\qquad\leq \prod_{k = 1}^{K}\left(y_k+\sigma_{\text{s}}^2\right)\cdot\prod_{n = 1}^{N-K}\sigma_{\text{s}}^2,
	\end{aligned}
\end{equation}
where $y_k$ is the $k$-th diagonal element of $\boldsymbol{S}^H\boldsymbol{R}_{\text{s}}\boldsymbol{S}$, which can be written as $\text{tr}\left[\boldsymbol{R}_{\text{s}}^{1/2}\left(\boldsymbol{s}_k\boldsymbol{s}_k^H\right)\boldsymbol{R}_{\text{s}}^{1/2}\right]$. Given the RIS beamforming matrix, the transmit beamforming problem can be formulated as
\begin{subequations}
	\label{eq16}
	\begin{align}
		\max_{\left\{\boldsymbol{s}_{k}\right\}} \quad & \sum_{k=1}^{K} \log \left(y_k+\sigma_{\text{s}}^2\right) \label{eq16a} \tag{17a} \\
		\text{s.t.} \quad &  R_{k} \geq R_{0}, \forall k \in \mathcal{K}, \label{eq16b} \tag{17b} \\
		&  \sum_{k=1}^{K} \operatorname{tr}\left(\boldsymbol{s}_{k}\boldsymbol{s}_{k}^H\right) \leq P_{0}, \label{eq16c} \tag{17c}
	\end{align}
\end{subequations}
where (\ref{eq16b}) is a non-convex constraint. We use the SDR method to handle (\ref{eq16a}). Define $ \boldsymbol{\Gamma}_k = \boldsymbol{s}_{k} \boldsymbol{s}_{k}^{H} $, and (\ref{eq10}) and (\ref{eq11}) are respectively reformulated as
\begin{equation}
	\label{eq17}
		\xi_{k}=\boldsymbol{\mu}_k^H\boldsymbol{\Gamma}_k\boldsymbol{\mu}_k,
\end{equation}
and
\begin{equation}
	\label{eq18}
	\zeta_{k} = \sum_{i=1,i \neq k}^{K}\boldsymbol{\mu}_k^H\boldsymbol{\Gamma}_i\boldsymbol{\mu}_k+ \sigma_{u_{k}}^{2},
\end{equation}
where 
\begin{equation}
	\label{eq18add1}
	\begin{aligned}
	\boldsymbol{\mu}_k =& \boldsymbol{h}_{\text{BU},k} +  	\boldsymbol{D\Theta}\boldsymbol{h}_{\text{RU},k}\\
	&+\sum_{l=1}^{L}\left(\boldsymbol{h}_{\text{BTU},l,k}+\boldsymbol{D\Theta h}_{\text{RTU},l,k}+\boldsymbol{H_{\text{BTR},l}\Theta h_{\text{RU},k}}\right).
	\end{aligned}
\end{equation}

Then, the transmit beamforming problem can be reformulated as
\begin{subequations}
	\label{eq19}
	\begin{align}
		\max_{\left\{\boldsymbol{\Gamma}_{k}\right\}} \quad & \sum_{k=1}^{K} \log 	\left(y_k+\sigma_{\text{s}}^2\right) \label{eq19a} \tag{20a} \\
		\text{s.t.} \quad &  R_{k} \geq R_{0}, \forall k \in \mathcal{K}, \label{eq19b} \tag{20b} \\
		&  \sum_{k=1}^{K} \operatorname{tr}\left(\boldsymbol{\Gamma}_{k}\right) \leq P_{0}, \label{eq19c} \tag{20c} \\
		& \operatorname{rank}\left(\boldsymbol{\Gamma}_{k}\right)=1, \forall k \in \mathcal{K}, \label{eq19d} \tag{20d}
	\end{align}
\end{subequations}
The problem in (\ref{eq19a}) can be transformed into a SDP problem by omitting the constraint in (\ref{eq19d}) constraint, which can be directly solved using CVX or other optimization solvers. \\
\textbf{\emph{Remark 3: }}\emph{The SDR method considered in this section is not always sufficient to obtain a rank-1 beamforming matrix for the problem (\ref{eq19a}). In order to achieve a rank-1 beamforming matrix, the Gaussian randomization method is used, which is a $\frac{\pi}{4}$-approximation of the optimal value \cite{add3}. It is worth noting that, unlike problem (\ref{eq16a}), where the optimization variables are complex vectors, problem (\ref{eq19a}) has matrices as optimization variables, resulting in additional computational burden for the SDR method. Although the SDR method is a classical approach for solving quadratically constrained quadratic programs (QCQP), it requires significant computing power when the number of optimization variables is large. The Gaussian randomization method is presented in \textbf{Algorithm 1}, where $G_{\text{max}}$ denotes the maximum number of Gaussian randomization iterations.}
\begin{algorithm}[t]
	\caption{Gaussian Randomization Method.}\label{alg:alg1}
	\small
	\begin{algorithmic}
		\STATE 
		\STATE 1. $ \textbf{Inputs: }$$\boldsymbol{\Gamma}_k$, $N_t$, $K$, $\boldsymbol{D}$, $\boldsymbol{H}_{\text{BTB},l}$, $\boldsymbol{H}_{\text{BTR},l}$, $\boldsymbol{H}_{\text{RTR},l}$, $\alpha_{k}$, $\boldsymbol{h}_{\text{BU},k}$, $\boldsymbol{h}_{\text{RU},k}$, $\sigma_{u_{k}}^2$, $\sigma_{u_{\text{s}}}^2$, $R_0$, $P_0$, $G_{\text{max}}$. 
		\STATE 2. $ \textbf{Outputs: }$Beamforming matrix $\boldsymbol{S}^*$ for (\ref{eq16a}).
		\FOR{$i=1$ to $G_{\text{max}}$}
		\FOR{$k=1$ to $K$}
		\STATE 3. Eignvalue decomposition $\boldsymbol{\Gamma}_k$, i.e., $\boldsymbol{\Gamma}_k = \boldsymbol{\Sigma}_{\text{c},k}\boldsymbol{V}_{\text{c},k}$, where $\boldsymbol{\Sigma}_{\text{c},k}$ is the eigenvalue matrix and $\boldsymbol{V}_{\text{c},k}$ is the eigenvector matrix.
		\STATE 4. Randomly generate $\tilde{\boldsymbol{x}}_k$, which is assumed to follow the standard CSCG distribution. Set $\tilde{\boldsymbol{s}}_k = \boldsymbol{V}_{\text{c},k}\boldsymbol{\Sigma}_{\text{c},k}^{1/2}\tilde{\boldsymbol{x}}_k $.
		\ENDFOR
		\STATE 5. Define $\tilde{\boldsymbol{S}}_i = \left(\tilde{\boldsymbol{s}}_1,\cdots\tilde{\boldsymbol{s}}_K\right)$.
		\IF{$\text{rank}(\tilde{\boldsymbol{S}}_i)=1$, and (\ref{eq19b}) and (\ref{eq19c}) are satisfied}
		\IF{$\tilde{\boldsymbol{S}}_i$ makes the radar MI bigger}
		\STATE 6. $\boldsymbol{S}^* =\tilde{\boldsymbol{S}}_i $.
		\ENDIF
		\ENDIF
		\ENDFOR
	\end{algorithmic}
	\label{alg1}
\end{algorithm}

\subsection{RIS Beamforming}
We now solve the RIS beamforming problem, assuming that the transmit beamforming is given and fixed. The RIS beamforming problem can be written as
\begin{subequations}
	\label{eq20}
\begin{align}
	\max_{\boldsymbol{\Theta}} \quad & \sum_{k=1}^{K} \log 	\left(y_k+\sigma_{\text{s}}^2\right) \label{eq20a} \tag{21a} \\
	\text{s.t.} \quad &  R_{k} \geq R_{0}, \forall k \in \mathcal{K}, \label{eq20b} \tag{21b} \\
	&  \boldsymbol{\Theta}_{m,m}\in\mathcal{F}, \forall m \in \mathcal{M}. \label{eq20c} \tag{21c}
\end{align}
\end{subequations}
The objective function in the formulated problem is a quartic function of $\boldsymbol{\Theta}$, and both (\ref{eq20b}) and (\ref{eq20c}) are non-convex. There exist no effective method to solve this problem, and it can only be solved through exhaustive search or genetic algorithms. We propose an efficient iterative method based on a one-dimensional search to optimize the phase shifts of the RIS elements. The basic idea is to iteratively optimize the phase shifts of $M$ RIS elements, gradually increasing the radar MI while satisfying the communication rate constraint. The optimal phase shift is selected from $\mathcal{F}$ for each RIS element. The proposed iterative method for optimizing the phase shifts of RIS elements is presented in \textbf{Algorithm 2}, where $ \delta^{\left(i\right)} $ denotes the objective value in each iteration, and $ \epsilon $ is a small threshold. At each iteration, the radar MI is non-decreasing, indicating that \textbf{Algorithm 2} can obtain the optimal RIS beamforming matrix. The proposed AO algorithm is presented in \textbf{Algorithm 3}. Additionally, $ \boldsymbol{S}^* $ and $ \boldsymbol{\Theta}^* $ represent the optimal DFRC BS beamforming matrix and the optimal RIS beamforming matrix, respectively.

\begin{algorithm}[t]
	\caption{Iterative Algorithm to Optimize the RIS Beamforming Matrix.}\label{alg:alg2}
	\small
	\begin{algorithmic}
		\STATE 
		\STATE 1. $ \textbf{Inputs: }$$N_t$, $N_r$, $M$, $K$, $L$, $\boldsymbol{S}$, $\boldsymbol{\Theta}$, $\boldsymbol{D}$, $\boldsymbol{D}^{\prime}$, $\boldsymbol{H}_{\text{BTB},l}$, $\boldsymbol{H}_{\text{BTR},l}$, $\boldsymbol{H}_{\text{RTR},l}$, $\alpha_{k}$, $\boldsymbol{h}_{\text{BU},k}$, $\boldsymbol{h}_{\text{RU},k}$, $\sigma_{\text{s}}^2$, $\sigma_{u_{k}}^2$, $R_0$, $i_{\text{max}}$, $\epsilon$. 
		\STATE 2. $ \textbf{Outputs: }$Beamforming matrices $\boldsymbol{\Theta}^*$ for (\ref{eq20a}).
		\WHILE {$i \leq i_{\text{max}}$, and $\delta^{\left(i\right)}-\delta^{\left(i-1\right)} \geq \epsilon$}
		\FOR{$m=1$ to $M$}
		\FOR{$\theta$ in $\mathcal{F}$}
		\STATE 3. Set the $m$-th diagonal element of $\boldsymbol{\Theta}^{\prime}$, i.e., $\boldsymbol{\Theta}_{m,m}^{\prime}$ to $\theta$.
		\IF{Constraints (\ref{eq20b}) are satisfied and the radar MI computed with $\boldsymbol{\Theta}^{\prime}$ increases}
		\STATE 4. Set $\boldsymbol{\Theta}_{m,m}$ to $\theta$.
		\ELSE
		\STATE 5. Keep $\boldsymbol{\Theta}_{m,m}$ fixed.
		\ENDIF
		\ENDFOR
		\ENDFOR
		\ENDWHILE
		\STATE 6. $\boldsymbol{\Theta}^* = \boldsymbol{\Theta}$.
	\end{algorithmic}
	\label{alg2}
\end{algorithm}

\begin{algorithm}[t]
	\caption{Alternating Optimization (AO) Algorithm to Maximize the Radar MI.}\label{alg:alg3}
	\small
	\begin{algorithmic}
		\STATE 
		\STATE 1. $ \textbf{Inputs: }$$N_t$, $N_r$, $M$, $K$, $L$, $\boldsymbol{S}$, $\boldsymbol{\Theta}$, $\boldsymbol{D}$, $\boldsymbol{D}^{\prime}$, $\boldsymbol{H}_{\text{BTB},l}$, $\boldsymbol{H}_{\text{BTR},l}$, $\boldsymbol{H}_{\text{RTR},l}$, $\alpha_{k}$, $\boldsymbol{h}_{\text{BU},k}$, $\boldsymbol{h}_{\text{RU},k}$, $\sigma_{\text{s}}^2$, $\sigma_{u_{k}}^2$, $R_0$, $i_{\text{max}}$, $\epsilon$. 
		\STATE 2. $ \textbf{Outputs: }$Beamforming matrices $\boldsymbol{S}^*$ and $\boldsymbol{\Theta}^*$ for (\ref{eq16a}) and (\ref{eq20a}), respectively.
		\STATE 3. $ \textbf{Initialization: }$ Randomly initialize $\boldsymbol{S}^{\left(1\right)} \!\!\text{, } \boldsymbol{\theta}^{\left(1\right)}\!\!$ that satisfy both constraints (\ref{eq13b}) and (\ref{eq13c}). Compute: $\delta^{\left(1\right)} \!\!$ according to (\ref{eq15}), set $ \delta^{\left(0\right)}\!\!=\!0,i\!=\!0 $.
		\WHILE {$i \leq i_{\text{max}}$, and $\delta^{\left(i\right)}-\delta^{\left(i-1\right)} \geq \epsilon$}
		\STATE 4. Fix $\boldsymbol{\theta}^{\left(i\right)}$, solve the convex problem (\ref{eq19a}).
		\IF {Constraint (\ref{eq19d}) is satisfied}
		\STATE 5. Obtain $ \boldsymbol{S}^{\left(i+1\right)} $ directly.
		\ELSE
		\STATE 6. Perform Gaussian randomization in \textbf{Algorithm 1} to obtain $ \boldsymbol{S}^{\left(i+1\right)} $.
		\ENDIF
		\STATE 7. Fix $\boldsymbol{S}^{\left(i+1\right)}$, using \textbf{Algorithm 2} to obtain $ \boldsymbol{\Theta}^{\left(i+1\right)} $.
		\STATE 8. $i=i+1$.
		\ENDWHILE
		\STATE 9. $\boldsymbol{S}^* = \boldsymbol{S}^{\left(i\right)}$, $\boldsymbol{\Theta}^* = \boldsymbol{\Theta}^{\left(i\right)}$.
	\end{algorithmic}
	\label{alg3}
\end{algorithm}

\section{Joint Beamforming on Manifold}
The SDR method proposed in Section \uppercase\expandafter{\romannumeral3} has a high computational complexity when the number of DFRC BS antennas or the number of RIS elements is large. We introduce a joint beamforming method on the manifold to address this issue. First, we formulate a weighted optimization problem where the transmit beamforming and the RIS beamforming are coupled. We transform the transmit and RIS beamforming to optimization problems on the complex hypersphere manifold and the complex circle manifold, respectively. Subsequently, we introduce an AO-RG algorithm to solve the weighted optimization problem iteratively. 
\subsection{Problem Formulation}
The weighted beamforming problem is expressed as
\begin{subequations}
	\label{eq24}
	\begin{align}
	\max_{\left\{\boldsymbol{s}_k\right\}, \boldsymbol{\theta}} \quad & \frac{\varepsilon I\left(\boldsymbol{Y}^{\text {s}} ; \boldsymbol{H}^{\text {s}} \mid \boldsymbol{S}\right)}{I_{\text{max}}}+\frac{(1-\varepsilon)\sum_{k=1}^{K}R_{k}}{R_{\text{max}}} \label{eq24a} \tag{22a} \\
	\text{s.t.} \quad &  \sum_{k=1}^{K} \operatorname{tr}\left(\boldsymbol{s}_{k}\boldsymbol{s}_{k}^H\right) \leq P_{0}, \label{eq24b} \tag{22b} \\
	& \boldsymbol{\theta}_{m}\in\mathcal{F}, \forall m \in\mathcal{M}, \label{eq24c} \tag{22c}
	\end{align}
\end{subequations}
where $I_{\text{max}}$ and $R_{\text{max}}$ are normalized factors, and $\varepsilon$ is the weighting factor. Furthermore, (\ref{eq24a}) is chosen to maximize the weighted radar MI and the communication rate; (\ref{eq24b}) is the transmit power constraint; (\ref{eq24c}) is the RIS phase shift constraint. Because the transmit beamforming and the RIS beamforming are coupled in (\ref{eq24a}), an AO-RG algorithm is proposed to solve (\ref{eq24a}).
\subsection{Transmit Beamforming on the Complex Hypersphere Manifold}
The first step of the proposed AO algorithm is to fix the RIS beamforming matrix. Then, the transmit beamforming problem can be formulated as 
\begin{equation}
	\max_{\boldsymbol{S}\in\mathcal{M}_1} \quad \frac{\varepsilon I\left(\boldsymbol{Y}^{\text {s}} ; \boldsymbol{H}^{\text {s}} \mid \boldsymbol{S}\right)}{I_{\text{max}}}+\frac{(1-\varepsilon)\sum_{k=1}^{K}R_{k}}{R_{\text{max}}}, \label{eq25}
\end{equation}
which is a problem on the complex hypersphere manifold $\mathcal{M}_1$, $ \mathcal{M}_1 = \{ \boldsymbol{S} \in \mathbb{C}^{N_t \times K}: \left\Vert \boldsymbol{S} \right\Vert_F=\sqrt{P_0} \} $, where $ \left\Vert \boldsymbol{S} \right\Vert_F = \sum_{k=1}^{K}\operatorname{tr}\left(\boldsymbol{s}_k\boldsymbol{s}_k^H\right) $. The constant terms of the radar MI are omitted in (\ref{eq25}). We solve problem (\ref{eq25}) using the Riemannian steepest ascent (RSA) algorithm. 

For any point $ \boldsymbol{S}\in \mathcal{M}_1 $, a \emph{tangent vector} at $ \boldsymbol{S} $ is defined as the vector that is tangential to any smooth curves on $ \mathcal{M}_1 $ through $ \boldsymbol{S} $. All the tangent vectors span the \emph{tangent space}. For the complex hypersphere manifold $ \mathcal{M}_1 $, the tangent space at $\boldsymbol{S}$ is defined as
\begin{equation}
	\label{eq26}
	T_{\boldsymbol{s}} \mathcal{M}_1 \triangleq \left\{\boldsymbol{Z} \in \mathbb{C}^{N_{t} \times K}: \operatorname{Re}\left\{\operatorname{tr}\left(\boldsymbol{S}^{H} \boldsymbol{Z}\right)\right\}=0\right\}.
\end{equation}
Let $ g\left(\boldsymbol{S}\right) \triangleq g_1\left(\boldsymbol{S}\right)+g_2\left(\boldsymbol{S}\right) $ be the objective function of problem (\ref{eq25}), with
\begin{equation}
	\label{eq27}
	g_{1}\left(\boldsymbol{S}\right)=\frac{\varepsilon}{I_{\text{max}}}\sum_{k=1}^{K} \log \left(y_k+\sigma_{\text{s}}^2\right),
\end{equation}
and 
\begin{equation}
	g_{2}(\boldsymbol{S})=\frac{(1-\varepsilon)}{R_{\text{max}}}\sum_{k=1}^{K}\alpha_{k} \log \left(1+\frac{\xi_{k}}{\zeta_{k}}\right). \label{eq28}
\end{equation}

The Euclidean gradient of $g\left(\boldsymbol{S}\right)$ is

\begin{equation}
	\label{eq29}
	\nabla_{\boldsymbol{s}} g=\left[\frac{\partial g_{1}}{\partial \boldsymbol{s}_{1}}+\frac{\partial g_{2}}{\partial \boldsymbol{s}_{1}}, \cdots, \frac{\partial g_{1}}{\partial \boldsymbol{s}_{k}}+\frac{\partial g_{2}}{\partial \boldsymbol{s}_{k}}, \cdots \frac{\partial g_{1}}{\partial \boldsymbol{s}_{K}}+\frac{\partial g_{2}}{\partial \boldsymbol{s}_{K}}\right],
\end{equation}
where
\begin{equation}
	\label{eq30}
	\frac{\partial g_{1}}{\partial \boldsymbol{s}_{k}}= \frac{2\varepsilon}{\ln 2\cdot I_{\text{max}}\left(y_k+\sigma_{\text{s}}^2\right)}  \boldsymbol{R}_{\text{s}}\boldsymbol{s}_k,
\end{equation}
and
\begin{equation}
	\label{eq31}
	\begin{aligned}
		&\frac{\partial g_{2}}{\partial \boldsymbol{s}_{k}}=\frac{2(1-\varepsilon)}{\ln 2\cdot R_{\text{max}}}\Bigg[\frac{\alpha_k\boldsymbol{\mu}_k^H\boldsymbol{s}_k\boldsymbol{\mu}_k}{\zeta_{k}\left(1+\frac{\xi_{k}}{\zeta_{k}}\right)}\\
		&-\sum_{i=1, i \neq k}^K\frac{\alpha_i\xi_{i}\boldsymbol{\mu}_i^H\boldsymbol{s}_k\boldsymbol{\mu}_i}{(\sigma_{u_{k}}^2+\boldsymbol{s}_k^H\boldsymbol{\mu}_i\boldsymbol{\mu}_i^H\boldsymbol{s}_k)^2\cdot\left(1+\frac{\xi_{i}}{\sigma_{u_{k}}^2+\boldsymbol{s}_k^H\boldsymbol{\mu}_i\boldsymbol{\mu}_i^H\boldsymbol{s}_k}\right)}\Bigg].
	\end{aligned}
\end{equation}

In order to derive the Riemannian gradient, we need to project the Euclidean gradient of problem (\ref{eq25}) onto the \emph{tangent space} (\ref{eq26}). The projection operator is defined as
\begin{equation}
	\label{eq33}
	\operatorname{grad}_{\boldsymbol{S}} g \triangleq \nabla_{\boldsymbol{S}} g-\operatorname{Re}\left\{\operatorname{tr}\left(\boldsymbol{S}^{H} \nabla_{\boldsymbol{s}} g\right)\right\} \boldsymbol{S}.
\end{equation}

In order to obtain the transmit beamforming matrix on $\mathcal{M}_1$, we define the iterative direction for the RSA. Assume that the direction of the $i$-th iteration for the RSA is $ \boldsymbol{\eta}_{i} $. Since the $i$-th iterative direction $\boldsymbol{\eta}_{i}$ and the $(i-1)$-th iterative direction $ \boldsymbol{\eta}_{i-1} $ are located in different \emph{tangent spaces}, i.e., $T_{\boldsymbol{S}^{(i)}}$ and $T_{\boldsymbol{S}^{(i-1)}}$, they can not be directly added. A projection operation is needed to perform the nonlinear combination. The $i$-th direction is defined as
\begin{equation}
	\label{eq34}
	\boldsymbol{\eta}_{i}\triangleq\operatorname{grad}_{\boldsymbol{S}^{(i)}} g+\delta_{i} T_{\boldsymbol{S}^{(i-1)} \rightarrow \boldsymbol{S}^{(i)}}\left(\boldsymbol{\eta}_{i-1}\right),
\end{equation}
where $T_{\boldsymbol{S}^{(i-1)} \rightarrow \boldsymbol{S}^{(i)}}\left(\boldsymbol{\eta}_{i-1}\right)$ is the projection operation that projects the $(i-1)$-th iterative direction $ \boldsymbol{\eta}_{i-1} $ onto the \emph{tangent space} of the $i$-th iteration $\boldsymbol{\eta}_{i}$, which is given as
\begin{equation}
	\label{eq35}
	\begin{aligned}
		T_{\boldsymbol{S}^{(i-1)} \rightarrow \boldsymbol{S}^{(i)}}\left(\boldsymbol{\eta}_{i-1}\right) &\triangleq T_{\boldsymbol{S}^{(i-1)}} \mathcal{M}_{1} \rightarrow T_{\boldsymbol{S}^{(i)}} \mathcal{M}_{1} \\
		&\triangleq\boldsymbol{\eta}_{i-1}-\operatorname{Re}\left\{\operatorname{tr}\left(\left(\boldsymbol{S}^{(i)}\right)^{H} \boldsymbol{\eta}_{i-1}\right)\right\} \boldsymbol{S}^{(i)},
	\end{aligned}
\end{equation}
and $\delta_{i}$ is the combination factor that is computed by the Polak-Ribi\.{e}re formula \cite{ref41}. The Polak-Ribi\.{e}re formula is defined by
\begin{equation}
	\label{eq36}
	\delta_{i}\triangleq\frac{\left\langle\operatorname{grad}_{\boldsymbol{S}^{(i)}} g, \boldsymbol{J}_{i}\right\rangle}{\left\langle\operatorname{grad}_{\boldsymbol{S}^{(i-1)}} g, \operatorname{grad}_{\boldsymbol{S}^{(i-1)}} g\right\rangle},
\end{equation}
where $\boldsymbol{J}_i $ is the difference between the gradient of the current iteration and the iterative direction of the last iteration, and $\boldsymbol{J}_i $ is defined as
\begin{equation}
	\label{eq37}
	\boldsymbol{J}_{i}\triangleq\operatorname{grad}_{\boldsymbol{S}^{(i)}} g-T_{\boldsymbol{S}^{(i-1)} \rightarrow \boldsymbol{S}^{(i)}}\left(\eta_{i-1}\right).
\end{equation}

Finally, the retraction projection is needed to project the points in the \emph{tangent space} onto the complex hypersphere manifold during each iteration of the RSA algorithm. The retraction projection operation is defined as
\begin{equation}
	\label{eq38}
	\mathcal{R}_{\boldsymbol{S}}\left(\boldsymbol{Z}\right) \triangleq \frac{\sqrt{P_{0}} \cdot(\boldsymbol{S}+\boldsymbol{Z})}{\left|\boldsymbol{S}+\boldsymbol{Z}\right|_{F}}.
\end{equation}

The RSA algorithm for solving problem (\ref{eq25}) is summarized in \textbf{Algorithm 4}.
\begin{algorithm}[t]
	\caption{Riemannian Steepest Ascent (RSA) Algorithm for Solving the Transmit Beamforming Subproblem.}\label{alg:alg4}
	\small
	\begin{algorithmic}
		\STATE 
		\STATE 1. $ \textbf{Inputs: }$$N_t$, $N_r$, $M$, $K$, $L$, $\boldsymbol{S}$, $\boldsymbol{\Theta}$, $\boldsymbol{R}_{\text{s}}$, $\boldsymbol{\mu}_{k}$, $\alpha_{k}$, $\sigma_{\text{s}}^2$, $\sigma_{u_{k}}^2$, $P_0$, $R_0$, $i_{\text{max}}$, $\epsilon$, $\varepsilon$. 
		\STATE 2. $ \textbf{Outputs: }$Beamforming matrices $\boldsymbol{S}^*$ for (\ref{eq25}).
		\STATE 3. $ \textbf{Initialization: }\!\!$ Randomly initialize $\boldsymbol{S}^{\left(0\right)}\!\!=\!\boldsymbol{S}^{\left(1\right)}\!\!\in \!\!\mathcal{M}_1$, compute $\operatorname{grad}_{\boldsymbol{S}^{(0)}}\!\! g $ and $\operatorname{grad}_{\boldsymbol{S}^{(1)}}\!\! g $ according to (\ref{eq29}), set $ \boldsymbol{\eta}_0\!\! =\! \operatorname{grad}_{\boldsymbol{S}^{(0)}} \!g$ and $i \!\!=\!\! 1$.
		\WHILE {$i \leq i_{\text{max}}$, and $\left\Vert \operatorname{grad}_{\boldsymbol{S}^{(i)}} g\right\Vert_F \geq \epsilon$}
		\STATE 4. Calculate the difference $\boldsymbol{J}_i $ according to (\ref{eq37}).
		\STATE 5. Calculate the combination factor $\delta_{i}$ according to (\ref{eq36}).
		\STATE 6. Calculate the search direction $\boldsymbol{\eta}_i$ according to (\ref{eq34}).
		\STATE 7. Compute $\boldsymbol{S}^{(i+1)}$ by the retraction projection operation (\ref{eq38}), i.e., $\boldsymbol{S}^{(i+1)} = \mathcal{R}_{\boldsymbol{S}^{(i)}}\left(\mu_i\boldsymbol{\eta}_i\right)$, where $\mu_i$ is the step size that is computed by line search method, such as Armijo rule.
		\STATE 8. $i=i+1$.
		\ENDWHILE
		\STATE 9. $\boldsymbol{S}^* = \boldsymbol{S}^{\left(i\right)}$.
	\end{algorithmic}
	\label{alg4}
\end{algorithm}

\subsection{RIS Beamforming on the Complex Circle Manifold}
Assuming the transmit beamforming matrix fixed, problem (\ref{eq24a}) can be reformulated as
\begin{equation}
	\max_{\boldsymbol{\theta}\in\mathcal{M}_2} \quad  \frac{\varepsilon I\left(\boldsymbol{Y}^{\text {s}} ; \boldsymbol{H}^{\text {s}} \mid \boldsymbol{S}\right)}{I_{\text{max}}}+\frac{(1-\varepsilon)\sum_{k=1}^{K}R_{k}}{R_{\text{max}}}, \label{eq39}
\end{equation}
which is an optimization problem on the complex circle manifold $\mathcal{M}_2$, $\mathcal{M}_2=\{\boldsymbol{\theta}\in\mathbb{C}^{M\times1}:\left\vert \theta_1\right\vert=\left\vert \theta_2\right\vert=\cdots=\left\vert \theta_M\right\vert=1\}$. In (\ref{eq39}), the discrete phase shift constraint is relaxed as a continuous one. Therefore, it is necessary to project the phase shifts of the RIS elements into the discrete set $\mathcal{F}$ after the optimal $\boldsymbol{\Theta}$ is obtained.

Similar to the transmit beamforming, we define some essential concepts for the complex circle manifold, which will be useful for solving the RIS beamforming by using the RSA algorithm. The \emph{tangent space} of any point $\boldsymbol{\theta}\in\mathcal{M}_2$ is given by
\begin{equation}
	\label{eq40}
	T_{\boldsymbol{\theta}} \mathcal{M}_{2}\triangleq\left\{\boldsymbol{z} \in \mathbb{C}^{M\time1}: \operatorname{Re}\left\{\boldsymbol{z} \circ \boldsymbol{\theta}^{H}\right\}=\boldsymbol{0}_{M}\right\}.
\end{equation}
The objective function of problem (\ref{eq39}) is $ g\left(\boldsymbol{\theta}\right) \triangleq g_1\left(\boldsymbol{\theta}\right)+g_2\left(\boldsymbol{\theta}\right) $. Then, the Euclidean gradient of $g(\boldsymbol{\theta})$ is given by $\nabla_{\boldsymbol{\theta}} g\left[\frac{\partial g_{1}(\boldsymbol{\theta})}{\partial \boldsymbol{\theta}}+\frac{\partial g_{2}(\boldsymbol{\theta})}{\partial \boldsymbol{\theta}}\right]$, where
\begin{equation}
	\label{eq41}
	\frac{\partial g_{1}}{\partial \boldsymbol{\theta}}= \frac{\varepsilon}{\ln 2\cdot I_{\text{max}}}\sum_{k=1}^{K}\frac{\partial\boldsymbol{s}_k^H\boldsymbol{R}_{\text{s}}\boldsymbol{s}/\partial \boldsymbol{\theta}}{y_k+\sigma_{\text{s}}^2},
\end{equation}
and
\begin{equation}
	\label{eq42}
	\frac{\partial g_{2}}{\partial \boldsymbol{\theta}}=\frac{(1-\varepsilon)}{\ln 2\cdot R_{\text{max}}}\sum_{k=1}^{K}\frac{\alpha_k}{1+\frac{\xi_k}{\zeta_k}}\left(\frac{1}{\zeta_k}\frac{\partial \xi_k}{\partial\boldsymbol{\theta}}-\frac{\xi_k}{\zeta_k^2}\frac{\partial \zeta_k}{\partial\boldsymbol{\theta}}\right).
\end{equation}
For simplicity, the partial derivatives $\frac{\partial\boldsymbol{s}_k^H\boldsymbol{R}_{\text{s}}\boldsymbol{s}}{\partial\boldsymbol{\theta}}$, $\frac{\partial \xi_k}{\partial\boldsymbol{\theta}}$, and $\frac{\partial \zeta_k}{\partial\boldsymbol{\theta}}$ are formulated in Appendix B. Then, to obtain the Riemannian gradient, the Euclidean gradient of $g(\boldsymbol{\theta})$ needs to be projected onto the \emph{tangent space} (\ref{eq40}) . The projection operator for the complex circle manifold is defined as
\begin{equation}
	\label{eq43}
	\operatorname{grad}_{\boldsymbol{\theta}} g \triangleq \nabla_{\boldsymbol{\theta}} g-\operatorname{Re}\left\{\nabla_{\boldsymbol{\theta}} g \circ \boldsymbol{\theta}^{H}\right\} \circ \boldsymbol{\theta}.
\end{equation}
Furthermore, the $i$-th iterative direction of the RSA algorithm is defined as 
\begin{equation}
	\label{eq44}
	\boldsymbol{\eta}_{i}=\operatorname{grad}_{\boldsymbol{\theta}^{(i)}} g+\delta_{i} T_{\boldsymbol{\theta}^{(i-1)} \rightarrow \boldsymbol{\theta}^{(i)}}\left(\boldsymbol{\eta}_{i-1}\right),
\end{equation}
where $T_{\boldsymbol{\theta}^{(i-1)} \rightarrow \boldsymbol{\theta}^{(i)}}\left(\boldsymbol{\eta}_{i-1}\right)$ and $\delta_{i}$ are the projection operation and the combination factor, respectively, and they can be expressed as
\begin{equation}
	\label{eq45}
	\begin{aligned}
		T_{\boldsymbol{\theta}^{(i-1)} \rightarrow \boldsymbol{\theta}^{(i)}}&\left(\boldsymbol{\eta}_{i-1}\right) \triangleq T_{\boldsymbol{\theta}^{(i-1)}} \mathcal{M}_{2} \rightarrow T_{\boldsymbol{\theta}^{(i)}} \mathcal{M}_{2} \\
		&\triangleq\boldsymbol{\eta}_{i-1}-\operatorname{Re}\left\{\boldsymbol{\eta}_{i-1}\circ\left(\boldsymbol{\theta}^{(i)}\right)^{H}\right\} \circ \boldsymbol{\theta}^{(i)}
	\end{aligned}
\end{equation}
and
\begin{equation}
	\label{eq46}
	\delta_{i}\triangleq\frac{\left\langle\operatorname{grad}_{\boldsymbol{\theta}^{(i)}} g, \boldsymbol{J}_{i}\right\rangle}{\left\langle\operatorname{grad}_{\boldsymbol{\theta}^{(i-1)}} g, \operatorname{grad}_{\boldsymbol{\theta}^{(i-1)}} g\right\rangle},
\end{equation}
where $ \boldsymbol{J}_i $ is the difference of directions, similar to that in (\ref{eq37}). Moreover, the retraction projection of the complex circle manifold is defined as
\begin{equation}
	\label{eq47}
	\mathcal{R}_{\boldsymbol{\theta}}\left(\boldsymbol{z}\right) \triangleq \left(\frac{z_1}{\vert z_1 \vert},\cdots, \frac{z_M}{\vert z_M \vert}\right)^T.
\end{equation}
The obtained continuous RIS phase vector is mapped to a discrete one, as follows:
\begin{equation}
	\label{eqadd2}
	\theta_m^*=\text{arg }\min_{\theta\in\mathcal{F}}\vert\theta_m-\theta\vert,\forall m \in\mathcal{M},
\end{equation}
where $\theta_m^*$ is the desired phase shift and $\theta_m$ is the continuous phase shift obtained by (\ref{eq47}).

The RSA algorithm for solving the RIS beamforming problem is summarized in \textbf{Algorithm 5}, and the proposed complete AO algorithm for solving the weighted optimization problem (\ref{eq24a}) is presented in \textbf{Algorithm 6}.
\begin{algorithm}[t]
	\caption{Riemannian Steepest Descent (RSA) Algorithm for Solving the RIS Beamforming Subproblem.}\label{alg:alg5}
	\small
	\begin{algorithmic}
		\STATE 
		\STATE 1. $ \textbf{Inputs: }$$N_t$, $N_r$, $M$, $K$, $L$, $\boldsymbol{S}$, $\boldsymbol{\Theta}$, $\boldsymbol{D}$, $\boldsymbol{D}^{\prime}$, $\boldsymbol{H}_{\text{BTB},l}$, $\boldsymbol{H}_{\text{BTR},l}$, $\boldsymbol{H}_{\text{RTR},l}$, $\alpha_{k}$, $\boldsymbol{h}_{\text{BU},k}$, $\boldsymbol{h}_{\text{RU},k}$, $\sigma_{\text{s}}^2$, $\sigma_{u_{k}}^2$, $R_0$, $i_{\text{max}}$, $\epsilon$, $\varepsilon$. 
		\STATE 2. $ \textbf{Outputs: }$Beamforming matrices $\boldsymbol{\theta}^*$ for (\ref{eq39}).
		\STATE 3. $ \textbf{Initialization: }$Randomly initialize $\boldsymbol{\theta}^{\left(0\right)}=\boldsymbol{\theta}^{\left(1\right)}\in \mathcal{M}_2$, compute $\operatorname{grad}_{\boldsymbol{\theta}^{(0)}} g $ and $\operatorname{grad}_{\boldsymbol{\theta}^{(1)}} g $ according to (\ref{eq43}), set $ \boldsymbol{\eta}_0 = \operatorname{grad}_{\boldsymbol{\theta}^{(0)}} g$ and $i = 1$.
		\WHILE {$i \leq i_{\text{max}}$, and $\left\Vert \operatorname{grad}_{\boldsymbol{\theta}^{(i)}} g\right\Vert_F \geq \epsilon$}
		\STATE 4. Calculate the difference $\boldsymbol{J}_i $.
		\STATE 5. Calculate the combination factor $\delta_{i}$ according to (\ref{eq46}).
		\STATE 6. Calculate the search direction $\boldsymbol{\eta}_i$ according to (\ref{eq44}).
		\STATE 7. Compute $\boldsymbol{\theta}^{(i+1)}$ by the retraction projection operation (\ref{eq47}), i.e., $\boldsymbol{\theta}^{(i+1)} = \mathcal{R}_{\boldsymbol{\theta}^{(i)}}\left(\mu_i\boldsymbol{\eta}_i\right)$, where $\mu_i$ is the step size that is compute by line search method, such as Armijo rule. \STATE \STATE 8. Project $\boldsymbol{\theta}^{(i+1)}$ as a discrete on by (\ref{eqadd2}).
		\STATE 9. $i=i+1$.
		\ENDWHILE
		\STATE 10. $\boldsymbol{\theta}^* = \boldsymbol{\theta}^{\left(i\right)}$.
	\end{algorithmic}
	\label{alg5}
\end{algorithm}
\begin{algorithm}[t]
	\caption{Alternating Optimization (AO) Algorithm for Solving the Weighted Optimization Problem (\ref{eq24a}).}\label{alg:alg6}
	\small
	\begin{algorithmic}
		\STATE 
		\STATE 1. $ \textbf{Inputs: }$$N_t$, $N_r$, $M$, $K$, $L$, $\boldsymbol{S}$, $\boldsymbol{\Theta}$, $\boldsymbol{D}$, $\boldsymbol{D}^{\prime}$, $\boldsymbol{H}_{\text{BTB},l}$, $\boldsymbol{H}_{\text{BTR},l}$, $\boldsymbol{H}_{\text{RTR},l}$, $\alpha_{k}$, $\boldsymbol{h}_{\text{BU},k}$, $\boldsymbol{h}_{\text{RU},k}$, $\sigma_{\text{s}}^2$, $\sigma_{u_{k}}^2$, $R_0$, $i_{\text{max}}$, $\epsilon$, $\varepsilon$.
		\STATE 2. $ \textbf{Outputs: }$Beamforming matrices $\boldsymbol{S}^*$ and $\boldsymbol{\theta}^*$ for (\ref{eq25}) and (\ref{eq39}), respectively.
		\STATE 3. $ \textbf{Initialization: }$ Apply the initialization of Algorithm 3 and Algorithm 4. Compute the objective value $\varsigma^{\left(1\right)} $ according to (\ref{eq24a}), set $ \varsigma^{\left(0\right)}=0$, $i=0$.
		\WHILE {$i \leq i_{\text{max}}$, and $\varsigma^{\left(i\right)}-\varsigma^{\left(i-1\right)} \geq \epsilon$}
		\STATE 4. Fix $\boldsymbol{\theta}^{\left(i\right)}$, and solve the complex hypersphere manifold optimization problem (\ref{eq25}) to obtain $ \boldsymbol{S}^{\left(i+1\right)} $.
		\STATE 5. Fix $\boldsymbol{S}^{\left(i+1\right)}$, and solve the complex circle manifold optimization problem (\ref{eq39}) to obtain $ \boldsymbol{\theta}^{\left(i+1\right)} $.
		\STATE 6. $i=i+1$.
		\ENDWHILE
		\STATE 7. $\boldsymbol{S}^* = \boldsymbol{S}^{\left(i\right)}$, $\boldsymbol{\theta}^* = \boldsymbol{\theta}^{\left(i\right)}$.
	\end{algorithmic}
	\label{alg6}
\end{algorithm}
\section{Complexity and Convergence Analysis}
We focus on analyzing the computational complexity of the SDR and ODI methods and the manifold optimization algorithms in each AO iteration since they dominate the complexity. However, predicting the total required iterations is challenging. The overall complexity of the two AO will be evaluated through simulations in the next section. The computational complexity is measured in flops \cite{ref42}.

\subsection{Complexity Analysis of the SDR Method and ODI Algorithm}
The SDR problems (\ref{eq19a}) can be solved with a worst-case complexity of $\mathcal{O}_{\text{SDR}}^{\text{Transmit}} = \mathcal{O}(N_t^4K^{0.5} + K^{4.5})$ \cite{ref43}. The overall computational complexity for the one-dimension iterative algorithm (\textbf{Algorithm} 2) is $\mathcal{O}_{\text{ODI}}^{\text{RIS}} =dM(40N_tM^2L+16N_tMN_rL+16N_t^2N_rL+16N_t^3L+8N_t^2ML+8N_t^2K+8M^2K+8N_tMK+32N_tK)$, where $d$ is the number of discrete phase shifts for each element of the RIS.
\subsection{Complexity Analysis of the RSA Algorithm for Transmit Beamforming}
The number of iterations required for the RSA algorithm (\textbf{Algorithm} 4) to converge is generally in the range of 5 to 10. We analyze the complexity for each iteration. As shown in line 4 in the RSA algorithm, the computations in each iteration are mainly due to the Euclidean gradient (\ref{eq29}), the Riemannian gradient (\ref{eq33}), and the projection operation (\ref{eq35}). Their orders of flops can be computed as $\mathcal{O}_{\text{RSA1}}^{\text{EG}} =24N_tK$, $\mathcal{O}_{\text{RSA1}}^{\text{RG}} = 12N_tK$, and $\mathcal{O}_{\text{RSA1}}^{\text{PR}} = 12N_tK$, respectively. Hence, the number of flops for calculating line 4 is $\mathcal{O}_{\text{RSA1}}^{\text{Line4}} = \mathcal{O}_{\text{RSA1}}^{\text{EG}} +\mathcal{O}_{\text{RSA1}}^{\text{RG}}+\mathcal{O}_{\text{RSA1}}^{\text{PR}}=48N_tK$. The complexity for calculating line 5, 6 ,7 are $16N_tK$, $4N_tK$, and $12N_tK$ flops, respectively. Therefore, the total complexity of the RSA algorithm is $\mathcal{O}_{\text{RSA1}}=\mathcal{O}\left(N_tK\right)$.
\subsection{Complexity Analysis of the RSA Algorithm for the RIS Beamforming}
Similarly, the iterations required for the RSA algorithm (\textbf{Algorithm} 5) to converge is in the range of 2 to 10. Here we focus on the complexity of each iteration. The number of flops for the Euclidean gradient (\ref{eq41}) and (\ref{eq42}), the Riemannian gradient (\ref{eq43}), and the projection operation (\ref{eq45}) are given by $\mathcal{O}_{\text{RSA2}}^{\text{EG}} = KL(64M^3+8N_t^2M+16N_tM^2+8N_rM^2+48M^2+16N_tM+6M)+K^2(24N_tM+16M^2+16N_t+8M)$, $\mathcal{O}_{\text{RSA2}}^{\text{RG}} = 12M$, and $\mathcal{O}_{\text{RSA2}}^{\text{PR}} = 12M$, respectively. Hence, the number of flops for line 4 in \textbf{Algorithm} 5 is $\mathcal{O}_{\text{RSA2}}^{\text{Line4}} = \mathcal{O}_{\text{RSA2}}^{\text{EG}} + \mathcal{O}_{\text{RSA2}}^{\text{RG}} + \mathcal{O}_{\text{RSA2}}^{\text{PR}}=KL(64M^3+8N_t^2M+16N_tM^2+8N_rM^2+48M^2+16N_tM+6M)+K^2(24N_tM+16M^2+16N_t+8M)+24M$. The complexity for lines 5, 6 ,7 are $16M$, $4M$, and $6M$ flops, respectively. Therefore, the complexity of the RSA algorithm is $\mathcal{O}_{\text{RSA2}}=\mathcal{O}\left(KL(M^3+N_t^2M+N_tM^2+N_rM^2)\right)$.
\subsection{Convergence Analysis}
In order to ensure the convergence of the AO algorithms, the sub-problem for updating each variable needs to be solved optimally in each iteration \cite{ref44}. The SDR method is proposed to solve the transmit beamforming and the RIS beamforming for maximizing the radar MI. However, the SDR method may not always be optimal since the Gaussian randomization induces an extra error. Nevertheless, the simulation results in Section \uppercase\expandafter{\romannumeral6} demonstrate that the beamforming matrices obtained by the SDR satisfy the rank-1 constraint in most cases, indicating that the proposed AO-SDR-ODI algorithm is convergent. Furthermore, the convergence of the proposed AO-RG algorithm will be confirmed by using simulation results in the next section.

\begin{figure}[htbp]
	\centering  %图片全局居中
	\subfigure[Radar MI versus the number of RIS elements.]{
		\includegraphics[width=0.8\linewidth]{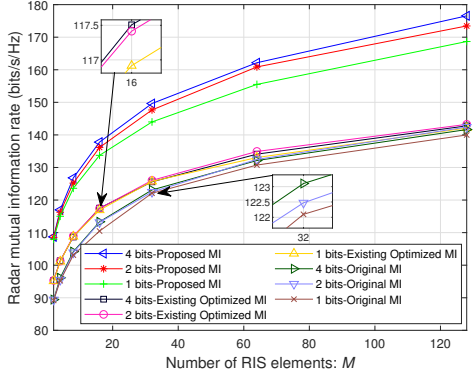}}
	\subfigure[MSE versus the number of RIS elements.]{
		\includegraphics[width=0.8\linewidth]{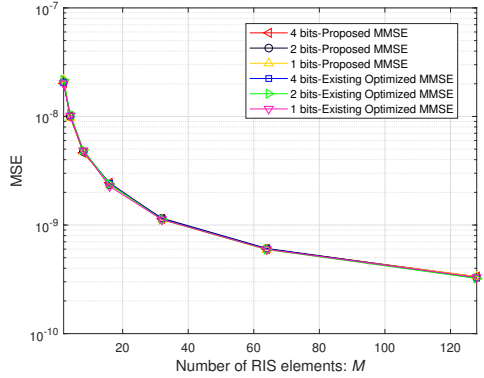}}
	\subfigure[Radar MI versus the transmit power budget.]{
		\includegraphics[width=0.8\linewidth]{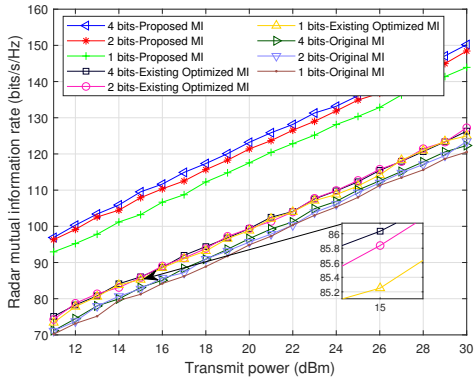}}
	\subfigure[Radar MI versus the communication rate threshold.]{
		\includegraphics[width=0.8\linewidth]{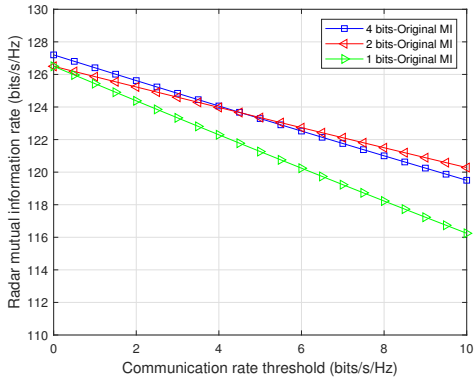}}
	\caption{Radar MI (a) and MSE (b) versus the number of RIS elements when the transmit power budget is 30 dBm. Radar MI versus the transmit power budget (c) and the communication rate threshold (d) when the number of RIS elements is 32. The threshold of the user rate $R_0$ in (c) is 4 bits/s/Hz. The transmit power in (d) is 30 dBm.}
	\label{fig2}
\end{figure}

\section{Simulation Results}
This section provides numerous simulation results to verify the effectiveness of the proposed algorithms. We assume that the BS is located at (0m, 0m); the RIS is located at (5m, 1m); the $K$ UEs and $L$ scatterers are uniformly distributed in a circle with a radius of 5 meters centered at (70m, 0m) and (90m, 10m), respectively. If not specified otherwise, the following simulations consider a DFRC system with 4 transmit and receive antennas, 2 scatterers, and 2 users, where the weighting factors for the user rates are both 2, the weighting factor $\varepsilon$ is 0.1, and the noise power is -110 dBm.
\subsection{Results for the Proposed AO-SDR-ODI Algorithm}
In order to verify the effectiveness of the proposed algorithms, the radar MI obtained by Eq. (18) in \cite{ref39} is utilized for comparison, which is labelled as ``Existing Optimized MI'' in the figures. We also plot the radar MI obtained by (\ref{eq7}) and (\ref{eq14}), which is labelled as ''Original MI''. The ``Proposed MI'' denotes the radar MI obtained by (\ref{eq7}) and (\ref{eq16}a). Moreover, the original MI coincides with the existing optimized MI when the communication rate threshold is $R_0=0$.

\begin{figure}[!t]
	\centering  %图片全局居中
	\subfigure[Radar MI for different signal paths versus the number of RIS elements.]{
		\includegraphics[width=0.8\linewidth]{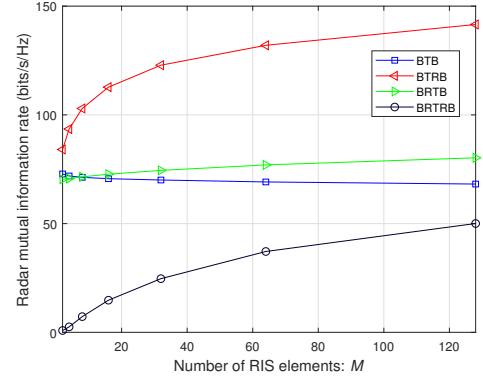}}
	\subfigure[Communication SINR for different signal paths versus the number of RIS elements.]{
		\includegraphics[width=0.8\linewidth]{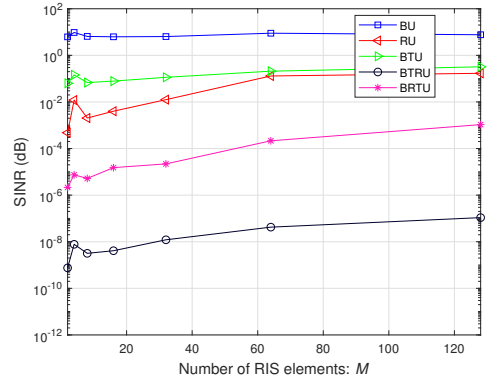}}
	\caption{Radar MI and communication SINR for different signal paths versus the number of RIS elements,when the transmit power budget is 30 dBm, the threshold of the user rate $R_0$ is 4 bits/s/Hz, and the RIS phase-shift quantization is 4 bits.}
	\label{fig3}
\end{figure}

\begin{figure}[!t]
	\centering  %图片全局居中
	\subfigure[Radar MI versus the number of RIS elements when the estimated channel parameters are imperfect.]{
		\includegraphics[width=0.85\linewidth]{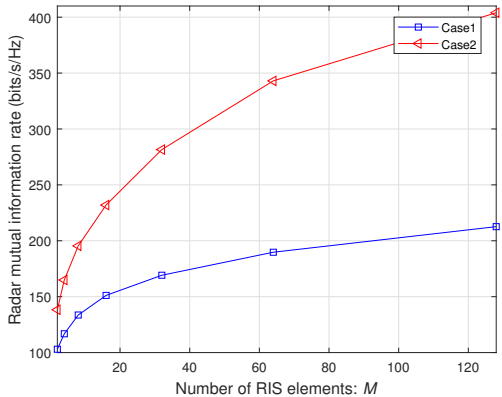}}
	\subfigure[Communication rate versus the number of RIS elements when the estimated channel parameters are imperfect.]{
		\includegraphics[width=0.85\linewidth]{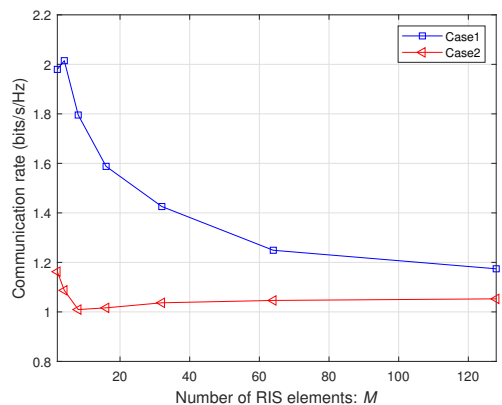}}
	\caption{Radar MI and communication rate versus the number of RIS elements, when the estimated channel parameters are imperfect. The transmit power budget is 30 dBm, the threshold of the user rate $R_0$ is 4 bits/s/Hz, and the RIS phase-shift quantization is 4 bits.}
	\label{fig4}
\end{figure}

\begin{figure}[!t]
	\centering  %图片全局居中
	\subfigure[Weighted information rate versus the number of RIS elements.]{
		\includegraphics[width=0.85\linewidth]{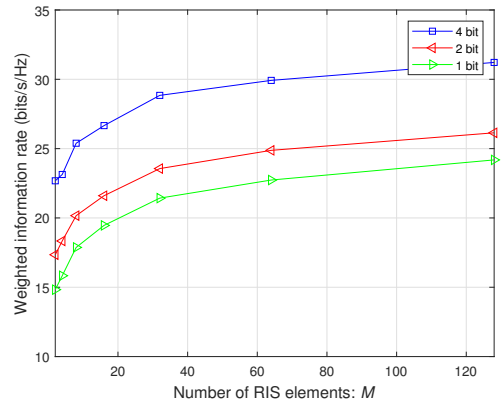}}
	\subfigure[Radar MI versus the number of RIS elements.]{
		\includegraphics[width=0.85\linewidth]{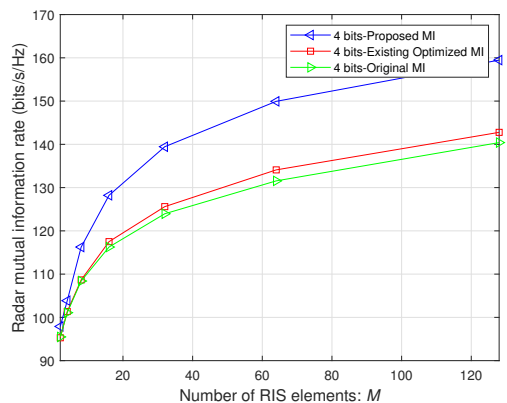}}
	\subfigure[Communication rate versus the number of RIS elements.]{
		\includegraphics[width=0.85\linewidth]{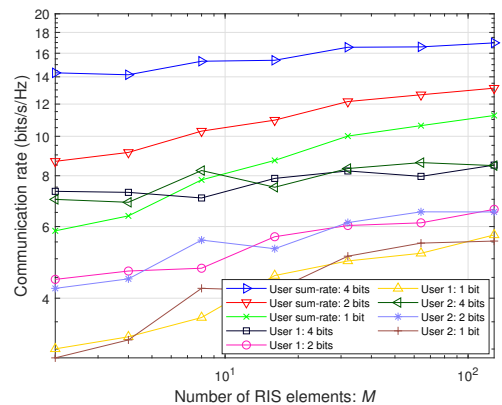}}
	\caption{Weighted information rate, radar MI, and communication rate versus the number of RIS elements, when the transmit power budget is 30 dBm.}
	\label{fig5}
\end{figure}

Figure \ref{fig2}(a) shows how the radar MI increases when increasing the number of RIS elements. The increasing trend is credited to the RIS providing extra signal paths, which enhance the strength of the radar echo. Increasing the number of bits for the
RIS phase shifts improves the radar MI. The results show that the proposed algorithm achieves a significantly better MI, compared to the original radar MI and the existing optimized radar MI, as it maximizes an upper bound of the original MI. Moreover, optimizing with respect to the upper bound effectively leads to the actual MI consistently increasing with the increasing of the bound, and the gap between them is small and stable. This validates the effectiveness of the proposed approach in maximizing the radar MI while satisfying the communication rate requirement. In addition, according to the obtained simulations, the SDR method can always obtain an optimal rank-1 transmit beamforming matrix, and the ODI method can obtain an optimal RIS beamforming matrix. Furthermore, the gap between the original and existing optimized radar MI is small and stable. This phenomenon is mainly due to the fact that there are overlapping signal paths for communication and sensing. These overlapping signal paths can provide mutual gains for communication and sensing.

Figure \ref{fig2}(b) shows that the sensing MSE of all schemes gradually decreases as the radar MI increases. The MSE values obtained by the proposed method and the existing optimized method, i.e., Eq. (100) in \cite{ref39}, are almost the same. This phenomenon indicates that the gap between the original radar MI and the existing optimized radar MI does not affect the MSE value. Moreover, the MSE values for different RIS phase shift quantization levels (i.e., 4-bit, 2-bit, and 1-bit) are almost the same. This indicates that the effect of the RIS phase-shift quantization levels on the MSE is negligible.

Figure \ref{fig2}(c) demonstrates that the radar MI steadily increases as the transmit power budget increases, as the feasible region of joint beamforming becomes larger when the transmit power budget increases. Similar to Fig. \ref{fig2}(a), the MI achieved by the proposed algorithm is greater than the existing optimized radar MI, which is further greater than the original radar MI. Moreover, increasing the number of bits for the RIS phase shifts improves the radar MI.

Figure \ref{fig2}(d) illustrates that the radar MI slowly decreases as the communication rate threshold increases, as the feasible region of joint beamforming becomes smaller when the communication rate threshold increases.

The performance for different signal paths is presented in Fig. \ref{fig3}(a) and (b). In Fig. \ref{fig3}(a), XTY, $\text{X,Y}\in\{\text{B,R}\}$, denotes the radar MI obtained by the proposed algorithm when only the signals from the XTY path is present. The SINRs for different signal paths are depicted in Fig. \ref{fig3}(b). We have two observations. Firstly, the BS-scatterers-RIS-BS and the BS-RIS-scatterers-BS signal paths have larger radar MI than the BS-RIS-scatterers-RIS-BS signal paths, as the latter signal path has a larger propagation loss than the former two signal paths; the BS-UEs paths have larger SINR than the BS-RIS-UEs and BS-scatterers-UE paths, and the above three paths, i.e., BS-UEs, BS-RIS-UEs, and BS-scatterers-UEs paths, have larger SINR than the BS-RIS-scatterers-UEs and BS-scatterers-RIS-UEs paths due to their larger path losses. Secondly, the gains of radar MI and communication SINR provided by the RIS-related signal paths become more significant, when increasing the number of RIS elements.

As obtaining perfect CSI in practical systems may be challenging, we simulate the performance of the proposed algorithm under CSI errors to verify its robustness. In Fig. \ref{fig4}(a) and (b), Case1 and Case2 represent 0.1 and 0.25 CSI error ratios, respectively. We define the CSI error ratio as $e$, and the CSI in the presence of errors is denoted as $\hat{\boldsymbol{H}} = \boldsymbol{H}+e\Vert\boldsymbol{H}\Vert_F$ or $\hat{\boldsymbol{h}} = \boldsymbol{h}+e\Vert\boldsymbol{h}\Vert_2$ according to \cite{ref40}. Figure \ref{fig4}(a) and (b) demonstrate that with an increasing proportion of CSI error ratios, the communication rate decreases while the radar MI increases. This phenomenon can be attributed to the fact that CSI errors relax the communication rate constraints, allowing for larger radar MI. Additionally, an increase in the number of RIS elements results in a higher channel dimension, amplifying the impact of CSI errors. Therefore, when the number of RIS elements is large and strict requirements on the communication and sensing performance are imposed, robust algorithms are needed to mitigate the impact of CSI errors, such as the algorithms in \cite{ref40,ref15}.

\begin{figure}[!t]
	\centering  %图片全局居中
	\subfigure[The cumulative distribution function (CDF) of the number of iterations.]{
		\includegraphics[width=0.85\linewidth]{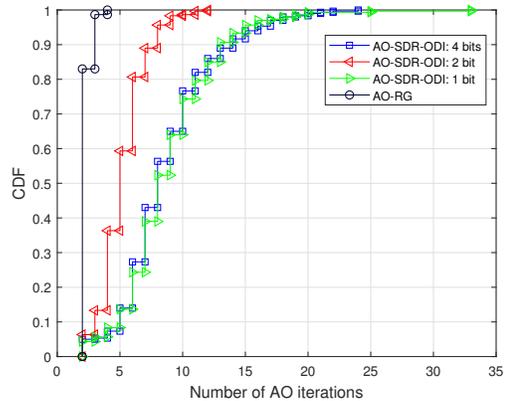}}
	\subfigure[The gradient norm of the proposed RSA algorithms in the first AO iteration.]{
		\includegraphics[width=0.85\linewidth]{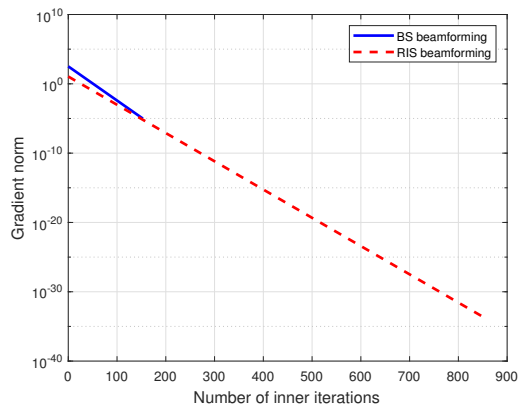}}
	\caption{The convergence performance of AO-SDR-ODI algorithm and AO-RG algorithm, when the number of RIS elements is 32.}
	\label{fig6}
\end{figure}

\subsection{Results for the Proposed AO-RG Algorithm}
In Fig. \ref{fig5}(a), it is shown that the weighted rate increases when increasing the number of RIS elements and the number of quantization levels of the RIS phase shifts. In Fig. \ref{fig5}(b), it is shown how the radar MI increases when increasing the RIS elements. The radar MI obtained by the proposed algorithm surpasses the original radar MI and the existing optimized radar MI. Since the radar MI for all RIS phase shift quantization levels is nearly identical, we only depict the 4 bits case. Moreover, optimizing with respect to the upper bound effectively leads to the actual MI consistently increasing with the increasing of the bound. These trends are aligned with the observation in Fig. \ref{fig2}(a). Figure 5(c) indicates that increasing the number of quantization levels leads to higher communication rates. This observation is consistent with the increase in the weighted rate and the radar MI observed in Fig. 5(a) and (b) as the RIS phase shift quantization levels increase.

Figure \ref{fig6}(a) and (b) show the convergence of the proposed AO-SDR-ODI and AO-RG algorithms. In Fig. \ref{fig6}(a), the AO-SDR-ODI algorithm converges within 35 AO iterations, while the AO-RG algorithm converges within 5 AO iterations. As depicted in Fig. \ref{fig6}(b), it is shown that within both BS beamforming and RIS beamforming, the gradient norm decreases rapidly with the number of inner loop iterations. Additionally, the weighted information acquired from each iteration of BS and RIS beamforming consistently demonstrates a non-decreasing pattern. As a result, a limited number of outer loop iterations is sufficient for the AO-RG algorithm to reach convergence. Although the AO-RG algorithm requires fewer iterations to converge, the initial values of the beamforming vectors need to be carefully chosen to avoid poor local optimal solutions. 

\section{Conclusion}
We have proposed a joint transmit and RIS beamforming design to guarantee small values of the MSE of sensing parameter estimation and communication rate. Two beamforming techniques, AO-SDR-ODI and AO-RG, are proposed to solve the joint beamforming problem. Specifically, the AO-SDR-ODI algorithm is proposed to maximize the radar MI under the communication rate constraints. In order to reduce the computational complexity, the AO-RG algorithm is proposed to solve the weighted joint beamforming problem. Simulation results demonstrate that both algorithms can achieve satisfactory communication and sensing performance. The AO-SDR-ODI algorithm results in a better accuracy at the cost of higher computational complexity. The AO-RG algorithm, on the other hand, has a lower computational complexity but relatively worse accuracy. Additionally, simulation results show that the MSE of sensing parameter estimation decreases as the radar MI increases.
\section*{Appendix A\\Derivation of the Differential Entropy}
The differential entropy $ h\left( \boldsymbol{Y}_{\text{s}} \vert \boldsymbol{SX} \right) $ can be formulated as
\begin{equation}
	\label{eq48}
	h\left( \boldsymbol{Y}_{\text{s}} \vert \boldsymbol{SX} \right) = \int -p\left( \boldsymbol{Y}_{\text{s}} \vert \boldsymbol{SX} \right)\log\left[ p\left( \boldsymbol{Y}_{\text{s}} \vert \boldsymbol{SX} \right) \right]d\boldsymbol{Y}_{\text{s}},
\end{equation}
where
\begin{equation}
	\label{eq49}
	\begin{aligned}
		&p\left( \boldsymbol{Y}_{\text{s}} \vert \boldsymbol{SX} \right) = \prod_{n=1}^{N_r}p\left[ \left( \boldsymbol{y}_{\text{s},n} \right)^T \vert \boldsymbol{SX} \right] \\
		& = \prod_{n=1}^{N_r}\frac{1}{\pi^{N}\det\left[ \sum_{l=1}^{L}\boldsymbol{\Lambda}_l + \boldsymbol{R}_W \right]}\\
		&\times \exp \left[ -\boldsymbol{y}_{\text{s},n}\left( \sum_{l=1}^{L}\boldsymbol{\Lambda}_l + \boldsymbol{R}_W \right)\left( \boldsymbol{y}_{\text{s},n} \right)^T \right] \\
		& = \frac{1}{\pi^{N_rN}\det^{N_r}\left[ \sum_{l=1}^{L}\boldsymbol{\Lambda}_l + \boldsymbol{R}_W \right]} \\
		&\times\exp\left\{ -\text{tr}\left[ \left( \sum_{l=1}^{L}\boldsymbol{\Lambda}_l + \boldsymbol{R}_W \right)^{-1}\boldsymbol{Y}_{\text{s}}\left(\boldsymbol{Y}_{\text{s}}\right)^H \right] \right\}.
	\end{aligned}
\end{equation}
Thus, we have
$h\left( \boldsymbol{Y}_{\text{s}} \vert \boldsymbol{SX} \right) = N_rN\log\left( \pi \right)+N_rN + N_r\log\left[ \det \left( \sum_{l=1}^{L}\boldsymbol{\Lambda}_l + \boldsymbol{R}_W \right) \right]$
The differential entropy $ h\left( \boldsymbol{W} \right) $ can be written as
\begin{equation}
	\label{eq51}
	h\left( \boldsymbol{W} \right) = \int -p\left( \boldsymbol{W} \right)\log\left[ p\left( \boldsymbol{W} \right) \right]d\boldsymbol{W},
\end{equation}
where
\begin{equation}
	\label{eq52}
	p\left( \boldsymbol{W} \right) = \frac{1}{\pi^{N_rN}\det^{N_r}\left( \boldsymbol{W} \right)}\exp\left\{ -\text{tr}\left[ \left( \boldsymbol{R}_W \right)^{-1}\boldsymbol{W}\boldsymbol{W}^H \right] \right\}.
\end{equation}
Then, the differential entropy $ h\left( \boldsymbol{W} \right) $ is
\begin{equation}
	\label{eq53}
	h\left( \boldsymbol{W} \right) = N_rN\log\left(\pi\right) +  N_rN + N_r\log\left[\det\left(\boldsymbol{R}_W\right)\right].
\end{equation}
\section*{Appendix B\\Derivation of the Partial Derivatives}
The partial derivative $\frac{\partial\boldsymbol{s}_k^H\boldsymbol{R}_{\text{s}}\boldsymbol{s}}{\partial\boldsymbol{\theta}}$ is expressed as
\begin{equation}
	\label{eq54}
	\begin{aligned}
		&\frac{\partial\boldsymbol{s}_k^H\boldsymbol{R}_{\text{s}}\boldsymbol{s}_k}{\partial\boldsymbol{\theta}} = \frac{\partial\boldsymbol{s}_k^H\sum_{l=1}^{L}\left(\hat{\boldsymbol{R}}_{\text{RTR},l}+\hat{\boldsymbol{R}}_{\text{BTR},l}+\hat{\boldsymbol{R}}_{\text{RTB},l}\right)\boldsymbol{s}_k}{\partial\boldsymbol{\theta}}\\
		&=\sum_{l=1}^{L}\Big\{\boldsymbol{\theta}^*\boldsymbol{\theta}^T\text{diag}(\boldsymbol{s}_k^H\boldsymbol{D})\boldsymbol{H}_{\text{RTR},l}\text{diag}(\boldsymbol{\theta})\boldsymbol{D}^{\prime}(\boldsymbol{D}^{\prime})^H\\
		&\cdot\text{diag}\left(\boldsymbol{H}_{\text{RTR},l}^H\text{diag}(\boldsymbol{D}^H\boldsymbol{s}_k)\right)+2\Big[\boldsymbol{\theta}^T\text{diag}(\boldsymbol{s}_k^H\boldsymbol{H}_{\text{BTR},l})\boldsymbol{D}^{\prime}(\boldsymbol{D}^{\prime})^H\\
		&\text{diag}(\boldsymbol{H}_{\text{BTR},l}^H\boldsymbol{s}_k)+\boldsymbol{\theta}^T\text{diag}(\boldsymbol{s}_k^H\boldsymbol{D})\boldsymbol{H}_{\text{RTB},l}\boldsymbol{H}_{\text{RTB},l}^H\text{diag}(\boldsymbol{D}^H\boldsymbol{s}_k)\Big]^T\Big\},
	\end{aligned}
\end{equation}
where the inner product Eq. (3.18) in \cite{refmanopt} is used. Moreover, the partial derivatives $\frac{\partial \xi_k}{\partial\boldsymbol{\theta}}$ and $\frac{\partial \zeta_k}{\partial\boldsymbol{\theta}}$ are formulated as
\begin{equation}
	\label{eq55}
	\begin{aligned}
	&\frac{\partial \xi_k}{\partial\boldsymbol{\theta}} =2\bigg[\text{diag}(\boldsymbol{h}_{\text{RU},k})^H\boldsymbol{D}^H\boldsymbol{s}_k\boldsymbol{s}_k^H\boldsymbol{h}_{\text{BU},k} \\
	&+ \text{diag}(\boldsymbol{h}_{\text{RU},k})^H\boldsymbol{D}^H\boldsymbol{s}_k\boldsymbol{s}_k^H\boldsymbol{D}\text{diag}(\boldsymbol{h}_{\text{RU},k})\boldsymbol{\theta}\bigg]\\
	&+\sum_{l_1=1}^{L}\Big\{2\big[\text{diag}(\boldsymbol{h}_{\text{RU},k})^H\boldsymbol{H}_{\text{BTR},l_1}^H\boldsymbol{s}_k\boldsymbol{s}_k^H\boldsymbol{h}_{\text{BU},k}\\
	&+\text{diag}(\boldsymbol{h}_{\text{RTU},l_1,k})\boldsymbol{D}^H\boldsymbol{s}_k\boldsymbol{s}_k^H\boldsymbol{h}_{\text{BU},k}\\
	&+\text{diag}(\boldsymbol{h}_{\text{RTU},l_1,k})\boldsymbol{D}^H\boldsymbol{s}_k\boldsymbol{s}_k^H\boldsymbol{D}\text{diag}(\boldsymbol{h}_{\text{RU},k})\boldsymbol{\theta}\\
	&+\text{diag}(\boldsymbol{h}_{\text{RU},k})^H\boldsymbol{H}_{\text{BTR},l_1}^H\boldsymbol{s}_k\boldsymbol{s}_k^H\boldsymbol{D}\text{diag}(\boldsymbol{h}_{\text{RU},k})\boldsymbol{\theta}\\
	&+\text{diag}(\boldsymbol{h}_{\text{RU},k})^H\boldsymbol{D}^H\boldsymbol{s}_k\boldsymbol{s}_k^H\boldsymbol{h}_{\text{BTU},l_1,k} \\
	&+ \text{diag}(\boldsymbol{h}_{\text{RU},k})^H\boldsymbol{D}^H\boldsymbol{s}_k\boldsymbol{s}_k^H\boldsymbol{D}\text{diag}(\boldsymbol{h}_{\text{RTU},l_1,k})\boldsymbol{\theta}\\
	&+\text{diag}(\boldsymbol{h}_{\text{RU},k})^H\boldsymbol{D}^H\boldsymbol{s}_k\boldsymbol{s}_k^H\boldsymbol{H}_{\text{BTR},l_1}\text{diag}(\boldsymbol{h}_{\text{RU},k})\boldsymbol{\theta}
	\big]\\
	&+\sum_{l_2=1}^{L}2\big[\text{diag}(\boldsymbol{h}_{\text{RTU},l_1,k})^H\boldsymbol{D}^H\boldsymbol{s}_k\boldsymbol{s}_k^H\boldsymbol{h}_{\text{BTU},l_2,k}\\
	&+\text{diag}(\boldsymbol{h}_{\text{RU},k})^H\boldsymbol{H}_{\text{BTR},l_1}^H\boldsymbol{s}_k\boldsymbol{s}_k^H\boldsymbol{h}_{\text{BTU},l_2,k} \\
	&+ \text{diag}(\boldsymbol{h}_{\text{RTU},l_1,k})^H\boldsymbol{D}^H\boldsymbol{s}_k\boldsymbol{s}_k^H\boldsymbol{D}\text{diag}(\boldsymbol{h}_{\text{RTU},l_2,k})\boldsymbol{\theta}\\
	&+\text{diag}(\boldsymbol{h}_{\text{RTU},l_1,k})^H\boldsymbol{D}^H\boldsymbol{s}_k\boldsymbol{s}_k^H\boldsymbol{H}_{\text{BTR},l_2,k}\text{diag}(\boldsymbol{h}_{\text{RU},k})\boldsymbol{\theta} \\
	&+ \text{diag}(\boldsymbol{h}_{\text{RU},k})^H\boldsymbol{H}_{\text{BTR},l_1}^H\boldsymbol{s}_k\boldsymbol{s}_k^H\boldsymbol{D}\text{diag}(\boldsymbol{h}_{\text{RTU},l_2,k})\boldsymbol{\theta}\\
	&+\text{diag}(\boldsymbol{h}_{\text{RU},k})^H\boldsymbol{H}_{\text{BTR},l_1}^H\boldsymbol{s}_k\boldsymbol{s}_k^H\boldsymbol{H}_{\text{BTR},l_2}\text{diag}(\boldsymbol{h}_{\text{RU},k})\boldsymbol{\theta} \big]\Big\},
\end{aligned}
\end{equation}
and
\begin{equation}
	\label{eq56}
	\begin{aligned}
	&\frac{\partial \zeta_k}{\partial\boldsymbol{\theta}} =\sum_{i=1,i \neq k}^{K}\bigg\{2\Big[\text{diag}(\boldsymbol{h}_{\text{RU},k})^H\boldsymbol{D}^H\boldsymbol{s}_i\boldsymbol{s}_i^H\boldsymbol{h}_{\text{BU},k} \\
	&+ \text{diag}(\boldsymbol{h}_{\text{RU},k})^H\boldsymbol{D}^H\boldsymbol{s}_i\boldsymbol{s}_i^H\boldsymbol{D}\text{diag}(\boldsymbol{h}_{\text{RU},k})\boldsymbol{\theta}\Big]\\
	&+\sum_{l_1=1}^{L}\Big\{2\big[\text{diag}(\boldsymbol{h}_{\text{RU},k})^H\boldsymbol{H}_{\text{BTR},l_1}^H\boldsymbol{s}_i\boldsymbol{s}_i^H\boldsymbol{h}_{\text{BU},k}\\
	&+\text{diag}(\boldsymbol{h}_{\text{RTU},l_1,k})\boldsymbol{D}^H\boldsymbol{s}_i\boldsymbol{s}_i^H\boldsymbol{h}_{\text{BU},k}\\
	&+\text{diag}(\boldsymbol{h}_{\text{RTU},l_1,k})\boldsymbol{D}^H\boldsymbol{s}_i\boldsymbol{s}_i^H\boldsymbol{D}\text{diag}(\boldsymbol{h}_{\text{RU},k})\boldsymbol{\theta}\\
	&+\text{diag}(\boldsymbol{h}_{\text{RU},k})^H\boldsymbol{H}_{\text{BTR},l_1}^H\boldsymbol{s}_i\boldsymbol{s}_i^H\boldsymbol{D}\text{diag}(\boldsymbol{h}_{\text{RU},k})\boldsymbol{\theta}\\
	&+\text{diag}(\boldsymbol{h}_{\text{RU},k})^H\boldsymbol{D}^H\boldsymbol{s}_i\boldsymbol{s}_i^H\boldsymbol{h}_{\text{BTU},l_1,k} \\
	&+ \text{diag}(\boldsymbol{h}_{\text{RU},k})^H\boldsymbol{D}^H\boldsymbol{s}_i\boldsymbol{s}_i^H\boldsymbol{D}\text{diag}(\boldsymbol{h}_{\text{RTU},l_1,k})\boldsymbol{\theta}\\
	&+\text{diag}(\boldsymbol{h}_{\text{RU},k})^H\boldsymbol{D}^H\boldsymbol{s}_i\boldsymbol{s}_i^H\boldsymbol{H}_{\text{BTR},l_1}\text{diag}(\boldsymbol{h}_{\text{RU},k})\boldsymbol{\theta}
	\big]\\
	&+\sum_{l_2=1}^{L}2\big[\text{diag}(\boldsymbol{h}_{\text{RTU},l_1,k})^H\boldsymbol{D}^H\boldsymbol{s}_i\boldsymbol{s}_i^H\boldsymbol{h}_{\text{BTU},l_2,k}\nonumber\\
	&+\text{diag}(\boldsymbol{h}_{\text{RU},k})^H\boldsymbol{H}_{\text{BTR},l_1}^H\boldsymbol{s}_i\boldsymbol{s}_i^H\boldsymbol{h}_{\text{BTU},l_2,k} \\
	&+ \text{diag}(\boldsymbol{h}_{\text{RTU},l_1,k})^H\boldsymbol{D}^H\boldsymbol{s}_i\boldsymbol{s}_i^H\boldsymbol{D}\text{diag}(\boldsymbol{h}_{\text{RTU},l_2,k})\boldsymbol{\theta}\\
	&+\text{diag}(\boldsymbol{h}_{\text{RTU},l_1,k})^H\boldsymbol{D}^H\boldsymbol{s}_i\boldsymbol{s}_i^H\boldsymbol{H}_{\text{BTR},l_2,k}\text{diag}(\boldsymbol{h}_{\text{RU},k})\boldsymbol{\theta} \\
	&+ \text{diag}(\boldsymbol{h}_{\text{RU},k})^H\boldsymbol{H}_{\text{BTR},l_1}^H\boldsymbol{s}_i\boldsymbol{s}_i^H\boldsymbol{D}\text{diag}(\boldsymbol{h}_{\text{RTU},l_2,k})\boldsymbol{\theta}\\
	&+\text{diag}(\boldsymbol{h}_{\text{RU},k})^H\boldsymbol{H}_{\text{BTR},l_1}^H\boldsymbol{s}_i\boldsymbol{s}_i^H\boldsymbol{H}_{\text{BTR},l_2}\text{diag}(\boldsymbol{h}_{\text{RU},k})\boldsymbol{\theta} \big]\Big\}\bigg\}.
\end{aligned}
\end{equation}

\begin{IEEEbiography}[{\includegraphics[width=1in,height=1.25in,clip,keepaspectratio]{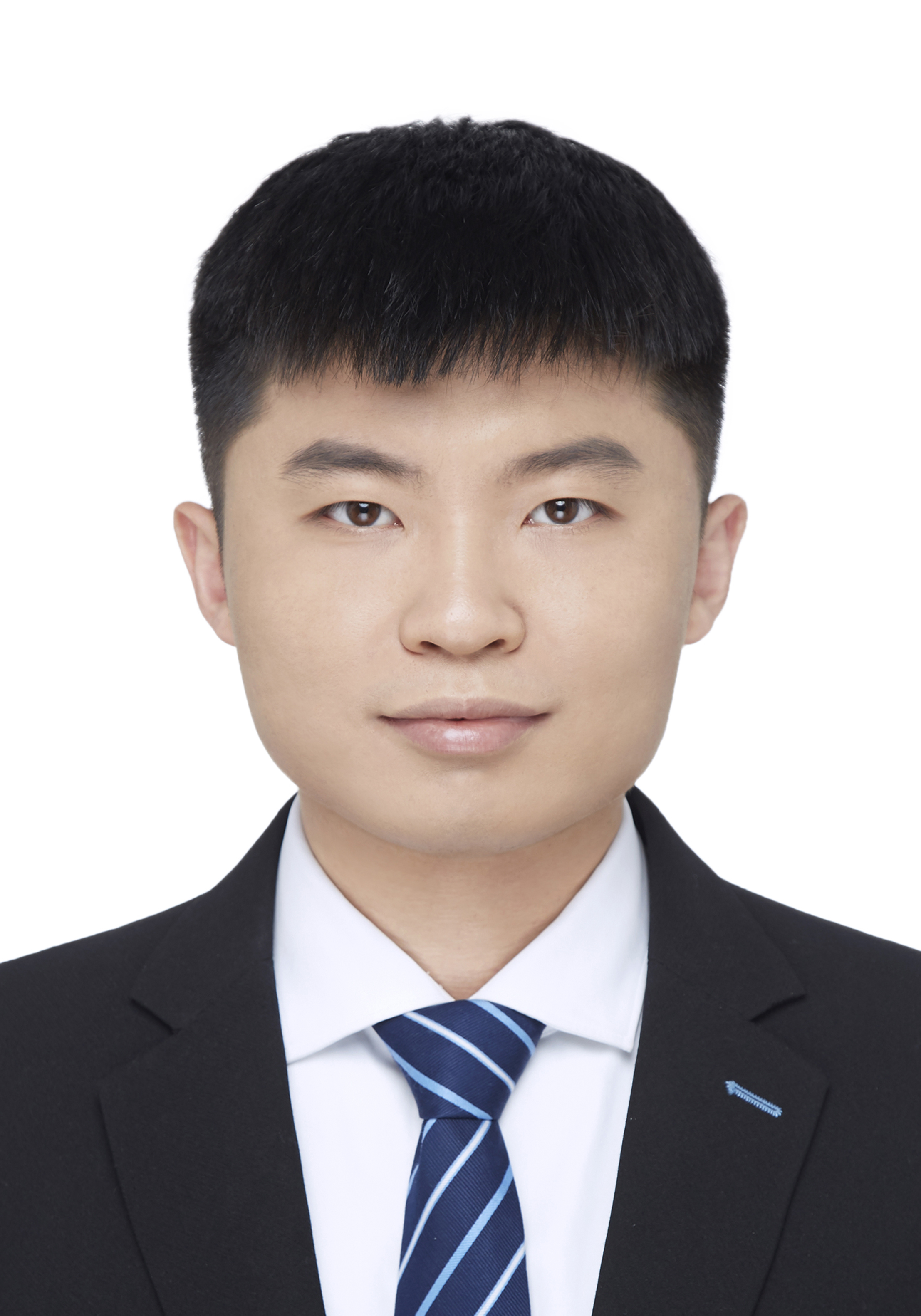}}]{Yongqing Xu}
	(Graduate Student Member, IEEE) received the bachelor's degree in communication engineering from the Xi'an University of Posts and Telecommunications (XUPT), Xi'an, China, in 2019. He is currently pursuing the Ph.D. degree with the School of Information and Communication Engineering of Beijing University of Posts and Telecommunications (BUPT), Beijing, China. His research mainly focuses on integrated sensing and communication (ISAC), reconfigurable intelligent surface (RIS), and the convex optimization.
\end{IEEEbiography}

\begin{IEEEbiography}[{\includegraphics[width=1in,height=1.25in,clip,keepaspectratio]{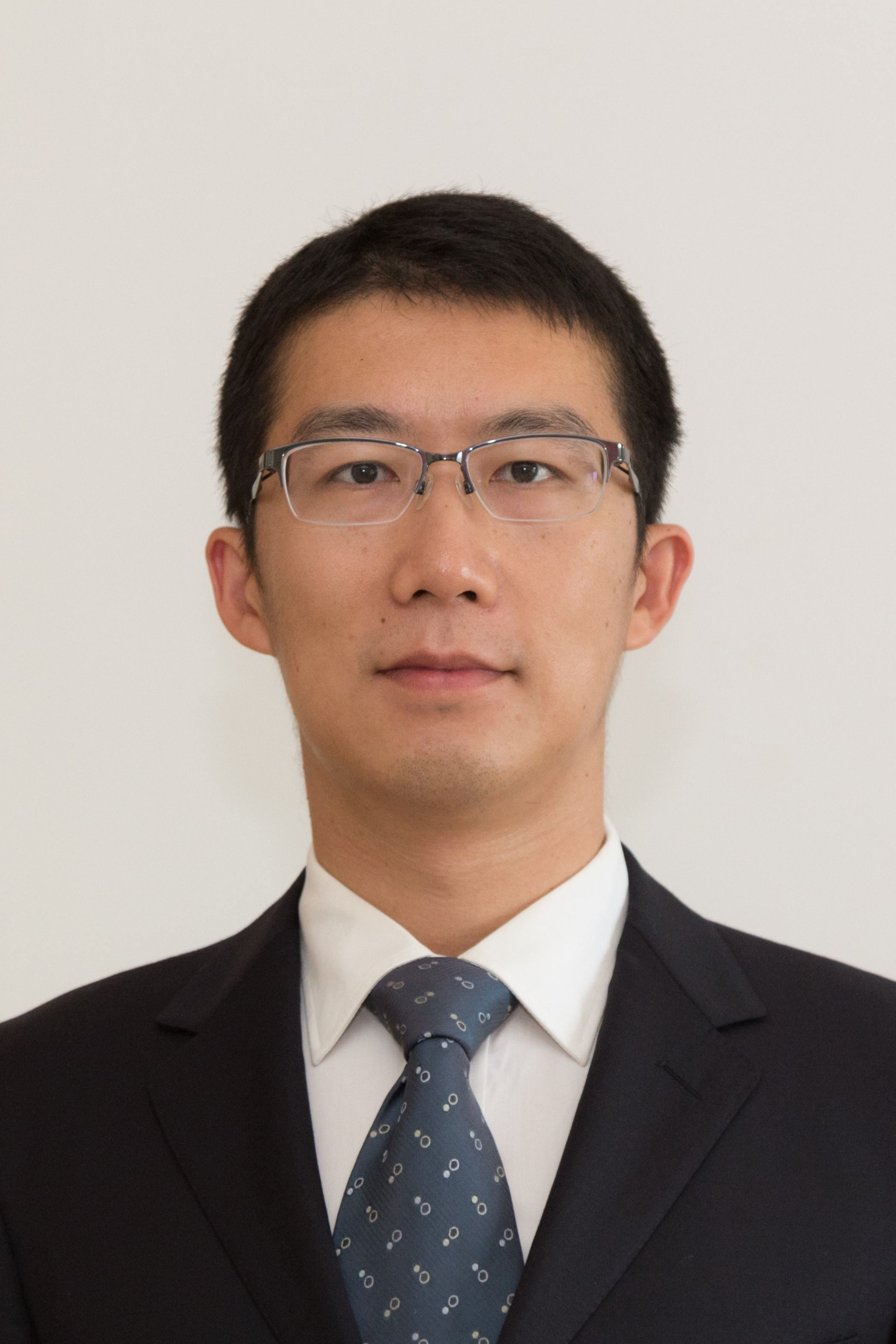}}]{Yong Li}
	(Member, IEEE) received the Ph.D. degree in signal and information processing from Beijing University of Posts and Telecommunications (BUPT), Beijing, China, in 2009.
	
	He is currently a Full Professor with the School of Information and Communication Engineering, BUPT. He has published more than 100 papers in journals, conference proceedings, and workshops. He holds more than 60 patents, including two U.S. patents. His current research interests include integrated sensing and communication, space-air-ground integrated networks, beyond-5G systems, and over-the-air (OTA) testing.
	
	Dr. Li was a recipient of the First Grade Award of Technological Invention from the China Institute of Communications.
\end{IEEEbiography}

\begin{IEEEbiography}[{\includegraphics[width=1in,height=1.25in,clip,keepaspectratio]{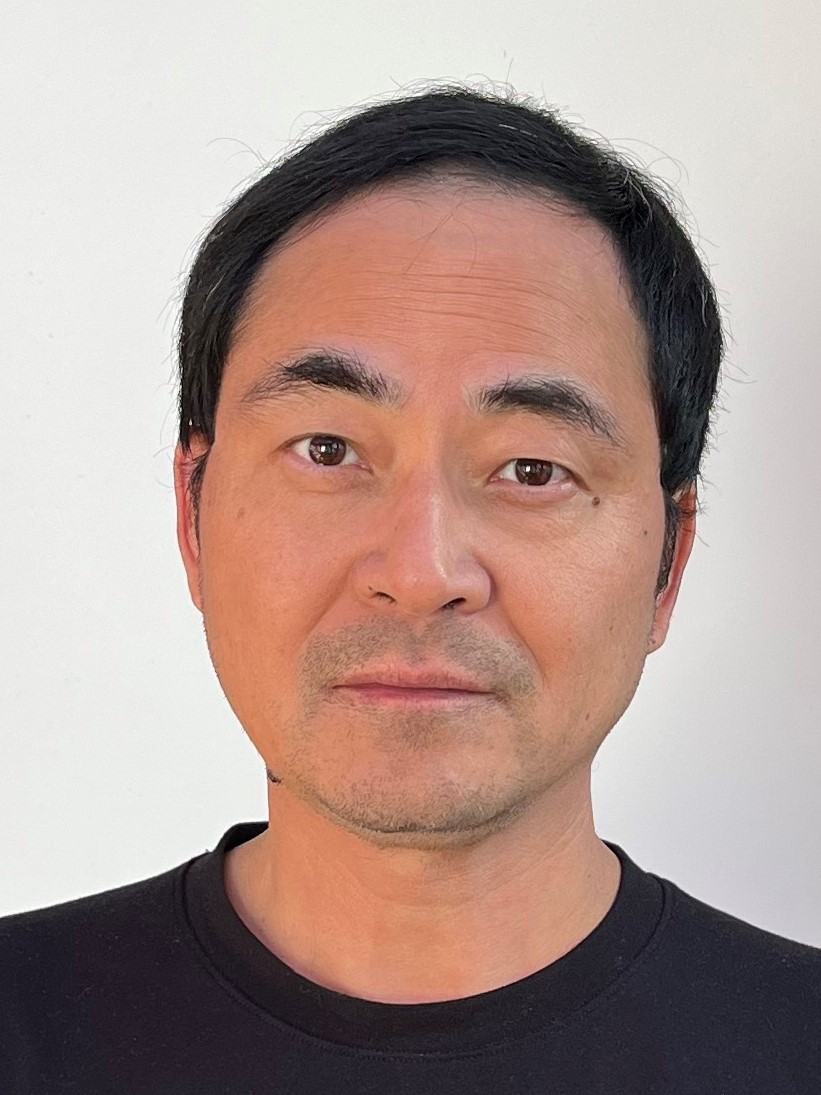}}]{J. Andrew Zhang}
	(M'04-SM'11) received the B.Sc. degree from Xi'an JiaoTong University, China, in 1996, the M.Sc. degree from Nanjing University of Posts and Telecommunications, China, in 1999, and the Ph.D. degree from the Australian National University, Australia, in 2004.
	
	Currently, Dr. Zhang is a Professor in the School of Electrical and Data Engineering, University of Technology Sydney, Australia. He was a researcher with Data61, CSIRO, Australia from 2010 to 2016, the Networked Systems, NICTA, Australia from 2004 to 2010, and ZTE Corp., Nanjing, China from 1999 to 2001.  Dr. Zhang's research interests are in the area of signal processing for wireless communications and sensing. He has published more than 270 papers in leading Journals and conference proceedings, and has won 5 best paper awards for his work including in IEEE ICC2013. He is a recipient of CSIRO Chairman's Medal and the Australian Engineering Innovation Award in 2012 for exceptional research achievements in multi-gigabit wireless communications.
\end{IEEEbiography}

\begin{IEEEbiography}[{\includegraphics[width=1in,clip,keepaspectratio]{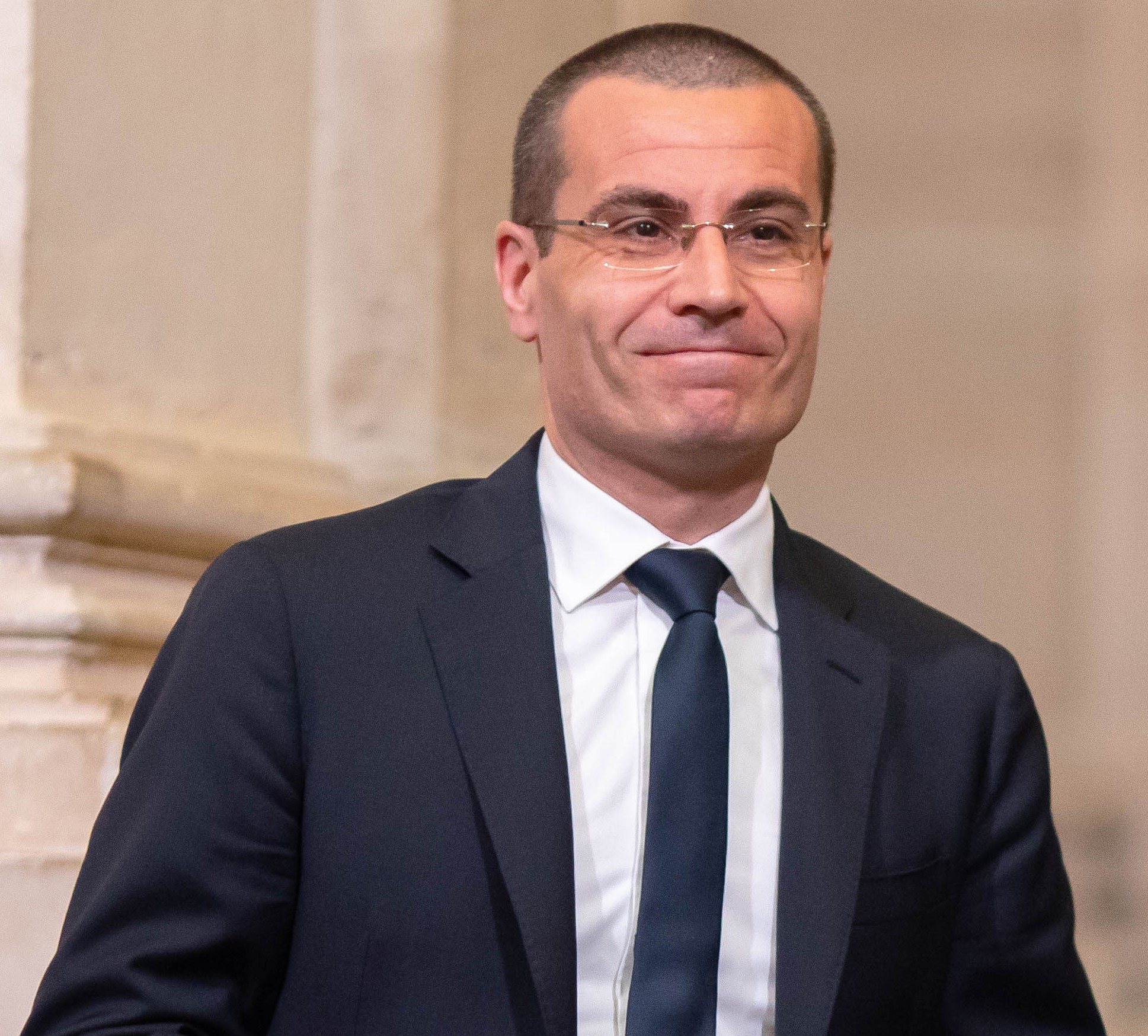}}]{Marco Di Renzo}
	(Fellow, IEEE) received the Laurea (cum laude) and Ph.D. degrees in electrical engineering from the University of L’Aquila, Italy, in 2003 and 2007, respectively, and the Habilitation à Diriger des Recherches (Doctor of Science) degree from University Paris-Sud (currently Paris-Saclay University), France, in 2013. Currently, he is a CNRS Research Director (Professor) and the Head of the Intelligent Physical Communications group in the Laboratory of Signals and Systems (L2S) at Paris-Saclay University – CNRS and CentraleSupelec, Paris, France. Also, he is an elected member of the L2S Board Council and a member of the L2S Management Committee, and is a Member of the Admission and Evaluation Committee of the Ph.D. School on Information and Communication Technologies, Paris-Saclay University. He is a Founding Member and the Academic Vice Chair of the Industry Specification Group (ISG) on Reconfigurable Intelligent Surfaces (RIS) within the European Telecommunications Standards Institute (ETSI), where he served as the Rapporteur for the work item on communication models, channel models, and evaluation methodologies. He is a Fellow of the IEEE, IET, and AAIA; an Ordinary Member of the European Academy of Sciences and Arts, an Ordinary Member of the Academia Europaea; and a Highly Cited Researcher. Also, he holds the 2023 France-Nokia Chair of Excellence in ICT, and was a Fulbright Fellow at City University of New York (USA), a Nokia Foundation Visiting Professor (Finland), and a Royal Academy of Engineering Distinguished Visiting Fellow (UK). His recent research awards include the 2021 EURASIP Best Paper Award, the 2022 IEEE COMSOC Outstanding Paper Award, the 2022 Michel Monpetit Prize conferred by the French Academy of Sciences, the 2023 EURASIP Best Paper Award, the 2023 IEEE ICC Best Paper Award, the 2023 IEEE COMSOC Fred W. Ellersick Prize, the 2023 IEEE COMSOC Heinrich Hertz Award, the 2023 IEEE VTS James Evans Avant Garde Award, and the 2023 IEEE COMSOC Technical Recognition Award from Signal Processing and Computing for Communications Technical Committee. He served as the Editor-in-Chief of IEEE Communications Letters during the period 2019-2023, and he is now serving in the Advisory Board. In 2024-2025, he will be serving as a Voting Member of the Fellow Evaluation Standing Committee and as the Director of Journals of the IEEE Communications Society.
\end{IEEEbiography}

\vspace{11pt}

\begin{IEEEbiography}[{\includegraphics[width=1in,height=1.25in,clip,keepaspectratio]{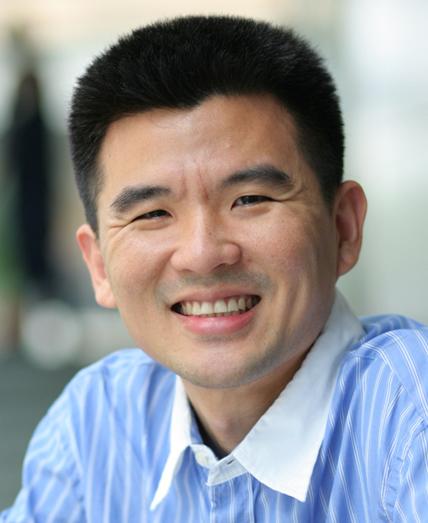}}]{Tony Q.S. Quek}
	(S'98-M'08-SM'12-F'18) received the B.E.\ and M.E.\ degrees in electrical and electronics engineering from the Tokyo Institute of Technology in 1998 and 2000, respectively, and the Ph.D.\ degree in electrical engineering and computer science from the Massachusetts Institute of Technology in 2008. Currently, he is the Cheng Tsang Man Chair Professor with Singapore University of Technology and Design (SUTD) and ST Engineering Distinguished Professor. He also serves as the Director of the Future Communications R\&D Programme, the Head of ISTD Pillar, and the Deputy Director of the SUTD-ZJU IDEA. His current research topics include wireless communications and networking, network intelligence, non-terrestrial networks, open radio access network, and 6G.
	
	Dr.\ Quek has been actively involved in organizing and chairing sessions, and has served as a member of the Technical Program Committee as well as symposium chairs in a number of international conferences. He is currently serving as an Area Editor for the {\scshape IEEE Transactions on Wireless Communications}. 
	
	Dr.\ Quek was honored with the 2008 Philip Yeo Prize for Outstanding Achievement in Research, the 2012 IEEE William R. Bennett Prize, the 2015 SUTD Outstanding Education Awards -- Excellence in Research, the 2016 IEEE Signal Processing Society Young Author Best Paper Award, the 2017 CTTC Early Achievement Award, the 2017 IEEE ComSoc AP Outstanding Paper Award, the 2020 IEEE Communications Society Young Author Best Paper Award, the 2020 IEEE Stephen O. Rice Prize, the 2020 Nokia Visiting Professor, and the 2022 IEEE Signal Processing Society Best Paper Award. He is a Fellow of IEEE and a Fellow of the Academy of Engineering Singapore.
\end{IEEEbiography}

\vfill


\begin{thebibliography}{99}
\bibliographystyle{IEEEtran}
\scriptsize
\bibitem{ref1}
F. Liu, C. Masouros, A. P. Petropulu, H. Griffiths, and L. Hanzo, ``Joint radar and communication design: Applications, state-of-the-art, and the road ahead,'' \emph{IEEE Trans. Wireless Commun.}, vol. 68, no. 6, pp. 3834–3862, Jun. 2020.
\bibitem{renew4}
E. C. Strinati {\it{et al}}., ``Wireless environment as a service enabled by reconfigurable intelligent surfaces: The RISE-6G perspective,'' in \emph{Proc. Joint EuCNC/6G Summit}, Porto, Portugal, Jun. 2021, pp. 562–567.
\bibitem{ref3}
J. A. Zhang {\it{et al}}., ``Enabling joint communication and radar sensing in mobile networks - a survey,'' \emph{IEEE Commun. Surveys Tuts.}, vol. 24, no. 1, pp. 306–345, Oct. 2021.
\bibitem{renew1}
H. Zhang {\it{et al}}., ``Holographic integrated sensing and communication,'' \emph{IEEE J. Sel. Areas Commun.}, vol. 40, no. 7, pp. 2114-2130, Jul. 2022.
\bibitem{ref4}
A. Babaei, W. H. Tranter, and T. Bose, ``A nullspace-based precoder with subspace expansion for radar/communications coexistence,'' in \emph{Proc. IEEE Glob. Commun. Conf. (GLOBECOM)}, Atlanta, GA, USA, Dec. 2013, pp. 3487–3492.
\bibitem{ref5}
F. Liu, C. Masouros, A. Li, and T. Ratnarajah, ``Robust MIMO beamforming for cellular and radar coexistence,'' \emph{IEEE Wireless Commun. Lett.}, vol. 6, no. 3, pp. 374–377, Jun. 2017.
\bibitem{ref6}
F. Liu, C. Masouros, A. Li, H. Sun, and L. Hanzo, ``MU-MIMO communications with MIMO radar: From co-existence to joint transmission,'' \emph{IEEE Trans. Wireless Commun.}, vol. 17, no. 4, pp. 2755–2770, Apr. 2018.
\bibitem{ref7}
T. Tian, T. Zhang, L. Kong, and Y. Deng, ``Transmit/Receive beamforming for MIMO-OFDM based dual-function radar and communication,'' \emph{IEEE Trans. Veh. Technol.}, vol. 70, no. 5, pp. 4693–4708, May 2021.
\bibitem{ref8}
Y. Liu, G. Liao, J. Xu, Z. Yang, and Y. Zhang, ``Adaptive OFDM integrated radar and communications waveform design based on information theory,'' \emph{IEEE Commun. Lett.}, vol. 21, no. 10, pp. 2174–2177, Oct. 2017.
\bibitem{add1}
Y. Yang and R. S. Blum, ``MIMO radar waveform design based on mutual information and minimum mean-square error estimation,'' \emph{IEEE Trans. Aerosp. Electron. Syst.}, vol. 43, no. 1, pp. 330–343, Jan. 2007.
\bibitem{add4}
M. R. Bell, ``Information theory and radar waveform design,'' \emph{IEEE Trans. Inf. Theory}, vol. 39, no. 5, pp. 1578-1597, Sept. 1993.
\bibitem{ref9}
K. Wu, J. A. Zhang, X. Huang, Y. J. Guo, and R. W. Heath, ``Waveform design and accurate channel estimation for frequency-hopping MIMO radar-based communications,'' \emph{IEEE Trans. Commun.}, vol. 69, no. 2, pp. 1244–1258, Feb. 2021.
\bibitem{ref10}
M. Di Renzo {\it{et al}}.,``Smart radio environments empowered by reconfigurable untelligent surfaces: How it works, state of research, and the road ahead,'' \emph{IEEE J. Sel. Areas Commun.}, vol. 38, no. 11, pp. 2450–2525, Nov. 2020.
\bibitem{ref11}
Y. Liu et al., ``Reconfigurable intelligent surfaces: Principles and opportunities,'' \emph{IEEE Commun. Surveys Tuts.}, vol. 23, no. 3, pp. 1546-1577, thirdquarter 2021.
\bibitem{renew5}
C. Pan et al., ``An overview of signal processing techniques for RIS/IRS-aided wireless systems,'' \emph{IEEE J. Sel. Topics Signal Process.}, vol. 16, no. 5, pp. 883-917, Aug. 2022.
\bibitem{ref12}
Q. Wu and R. Zhang, ``Beamforming optimization for wireless network aided by intelligent reflecting surface with discrete phase shifts,'' \emph{IEEE Trans. Commun.}, vol. 68, no. 3, pp. 1838–1851, Mar. 2020.
\bibitem{ref14}
X. Yu, D. Xu, and R. Schober, ``MISO wireless communication systems via intelligent reflecting surfaces: (invited paper),'' in \emph{Proc. IEEE/CIC Int. Conf. Commun. China (ICCC)}, ChangChun, China, Aug. 2019, pp. 735–740.
\bibitem{ref15}
G. Zhou, C. Pan, H. Ren, K. Wang, and A. Nallanathan, ``A framework of robust transmission design for IRS-aided MISO communications with imperfect cascaded channels,'' \emph{IEEE Trans. Signal Process.}, vol. 68, pp. 5092–5106, Aug. 2020.
	\bibitem{renew3}
	X. Qian, M. Di Renzo, J. Liu, A. Kammoun and M. S. Alouini, ``Beamforming through reconfigurable intelligent surfaces in single-user MIMO systems: SNR distribution and scaling laws in the presence of channel fading and phase noise,'' \emph{IEEE Wireless Commun. Lett.}, vol. 10, no. 1, pp. 77-81, Jan. 2021.
	\bibitem{ref16}
	Y.-C. Liang {\it{et al.}}, ``Reconfigurable intelligent surfaces for smart wireless environments: channel estimation, system design and applications in 6G networks,'' \emph{Sci. China Inf. Sci.}, vol. 64, no. 10, pp. 1–21, 2021.
	\bibitem{ref17}
	T. L. Jensen {\it{et al.}}, ``An optimal channel estimation scheme for intelligent reflecting surfaces based on a minimum variance unbiased estimator,'' in \emph{Proc. IEEE Int. Conf. on Acoust., Speech and Signal Process. (ICASSP)}, May 2020, pp. 5000–5004.
	\bibitem{ref19}
	J. He {\it{et al.}}, ``Channel Estimation for RIS-Aided mmWave MIMO Systems via Atomic Norm Minimization,'' \emph{IEEE Trans. Wireless Commun.}, vol. 20, no. 9, pp. 5786–5797, Sep. 2021.
	\bibitem{ref21}
	L. Wei {\it{et al.}}, ``Channel Estimation for RIS-Empowered Multi-User MISO Wireless Communications,'' \emph{IEEE Trans. Commun.}, vol. 69, no. 6, pp. 4144–4157, Jun. 2021.
	\bibitem{ref24}
	C. Liu {\it{et al.}}, ``Deep residual learning for channel estimation in intelligent reflecting surface-assisted multi-user communications,'' IEEE Trans. Wireless Commun., vol. 21, no. 2, pp. 898–912, Feb. 2022.
	\bibitem{ref25}
	A. Aubry, A. De Maio, and M. Rosamilia, ``Reconfigurable intelligent surfaces for N-LOS radar surveillance,'' \emph{IEEE Trans. Veh. Technol.}, vol. 70, no. 10, pp. 10735-10749, Oct. 2021.
	\bibitem{ref26}
	F. Wang, H. Li, and J. Fang, ``Joint active and passive beamforming for IRS-assisted radar,'' \emph{IEEE Signal Process. Lett.}, vol. 29, pp. 349–353, Dec. 2022.
	\bibitem{renew2}
	J. Hu {\it{et al.}}, ``MetaSensing: Intelligent metasurface assisted RF 3D sensing by deep reinforcement learning,'' \emph{IEEE J. Sel. Areas Commun.}, vol. 39, no. 7, pp. 2182-2197, Jul. 2021.
	\bibitem{ref27}
	X. Shao, C. You, W. Ma, X. Chen, and R. Zhang, ``Target sensing with intelligent reflecting surface: Architecture and performance,'' \emph{IEEE J. Sel. Areas Commun.}, 2022, doi: 10.1109/JSAC.2022.3155546.
	\bibitem{ref28}
	A. Elzanaty {\it{et al.}}, ``Reconfigurable intelligent surfaces for localization: Position and orientation error bounds,'' \emph{IEEE Trans. Signal Process.}, vol. 69, pp. 5386–5402, 2021.
	\bibitem{ref29}
	H. Zhang {\it{et al.}}, ``Towards ubiquitous positioning by leveraging reconfigurable intelligent surface,'' \emph{IEEE Commun. Lett.}, vol. 25, no. 1, pp. 284–288, Jan. 2021.
	\bibitem{ref30}
	E. Bj{\"o}rnson {\it{et al.}}, ``Reconfigurable intelligent surfaces: A signal processing perspective with wireless applications,'' \emph{IEEE Signal Process. Mag.}, vol. 39, no. 2, pp. 135–158, Mar. 2022.
	\bibitem{ref31}
	K. Keykhosravi {\it{et al.}}, ``SISO RIS-enabled joint 3D downlink localization and synchronization,'' in \emph{Proc. IEEE Int. Conf. Commun.}, Jun. 2021, pp. 1–6.
\bibitem{ref32}
X. Wang, Z. Fei, J. Guo, Z. Zheng, and B. Li, ``RIS-assisted spectrum sharing between MIMO radar and MU-MISO communication systems,'' \emph{IEEE Wireless Commun. Lett.}, vol. 10, no. 3, pp. 594–598, Mar. 2021.
\bibitem{ref33}
Y. He, Y. Cai, H. Mao, and G. Yu, ``RIS-assisted communication radar coexistence: Joint beamforming design and analysis,'' \emph{IEEE J. Sel. Areas Commun.}, 2022, doi: 10.1109/JSAC.2022.3155507.
\bibitem{ref34}
R. S. Prasobh Sankar, B. Deepak, and S. P. Chepuri, ``Joint communication and radar sensing with reconfigurable intelligent surfaces,'' in \emph{IEEE 22nd Int. Workshop Signal Process. Adv. Wireless Commun. (SPAWC)}, Piscataway, NJ, USA, Sep. 2021, pp. 471–475.
\bibitem{ref35}
Z.-M. Jiang {\it{et al}}., ``Intelligent reflecting surface aided dual-function radar and communication system,'' \emph{IEEE Syst. J.}, vol. 16, no. 1, pp. 475–486, Feb. 2021.
\bibitem{ref36}
X. Wang, Z. Fei, Z. Zheng, and J. Guo, ``Joint waveform design and passive beamforming for RIS-assisted dual-functional radar-communication system,'' \emph{IEEE Trans. Veh. Technol.}, vol. 70, no. 5, pp. 5131–5136, May 2021.
\bibitem{ref37}
X. Liu {\it{et al}}., ``Proximal policy optimization-based transmit beamforming and phase-shift design in an IRS-aided ISAC system for the THz band,'' \emph{IEEE J. Sel. Areas Commun.}, 2022, doi: 10.1109/JSAC.2022.3158696.
\bibitem{refHRIS1}
G. C. Alexandropoulos, N. Shlezinger, I. Alamzadeh, M. F. Imani, H. Zhang, and Y. C. Eldar,``Hybrid reconfigurable
intelligent metasurfaces: Enabling simultaneous tunable reflections and sensing for 6G wireless communications,'' \emph{arXiv preprint arXiv}: 2104.04690, 2021.
\bibitem{refHRIS2}
S. P. Chepuri, N. Shlezinger, Fan Liu, G. C. Alexandropoulos, S. Buzzi, and Y. C. Eldar, ``Integrated Sensing and Communications with Reconfigurable Intelligent Surfaces,'' \emph{arXiv preprint arXiv}: 2211.01003, 2022.
\bibitem{ref38}
F. Liu and C. Masouros, ``A tutorial on joint radar and communication transmission for vehicular networks—part III: Predictive beamforming without state models,'' \emph{IEEE Commun. Lett.}, vol. 25, no. 2, pp. 332–336, Feb. 2021.

\bibitem{reftimedelay}
F. Liu, Y.-F. Liu, A. Li, C. Masouros, and Y. C. Eldar, ``Cramér-Rao bound optimization for joint radar-communication beamforming,'' \emph{IEEE Trans. Signal Process.}, vol. 70, pp. 240–253, 2022.

\bibitem{zhang2022joint}
R. Liu, M. Li, Y. Liu, Q. Wu and Q. Liu, ``Joint transmit waveform and passive beamforming design for RIS-aided DFRC systems,'' \emph{IEEE J. Sel. Topics Signal Process.}, vol. 16, no. 5, pp. 995-1010, Aug. 2022.

\bibitem{ref39}
B. Tang, J. Tang, and Y. Peng, ``MIMO radar waveform design in colored noise based on information theory,'' \emph{IEEE Trans. Signal Process.}, vol. 58, no. 9, pp. 4684–4697, Sep. 2010.

\bibitem{ref40}
Y. Xu, Y. Li and J. Wu, ``Weighted sum-rate outage probability constrained transmission design for IRS-enhanced communication,'' \emph{in Proc. IEEE Wireless Commun. Networking Conf. (WCNC)}, Austin, TX, United states, Apr. 2022, pp. 1081-1086.
\bibitem{add3}
A. M.-C. So, J. Zhang, and Y. Ye, ``On approximating complex quadratic optimization problems via semidefinite programming relaxations,'' \emph{Mathematical Programming}, vol. 110, no. 1, pp. 93–110, Jun. 2007.

\bibitem{ref41}
J. Nocedal and S. J. Wright, \emph{Numerical Optimization}. NewYork, NY, USA: Springer, 2006.

\bibitem{ref42}
S. Boyd and L. Vandenberghe, Convex Optimization. Cambridge, U.K.: Cambridge Univ. Press, 2004.
\bibitem{ref43}
Z. Luo, W. Ma, A. M. So, Y. Ye, and S. Zhang, ``Semidefinite relaxation of quadratic optimization problems,'' \emph{IEEE Signal Process. Mag.}, vol. 27, no. 3, pp. 20–34, 2010.
\bibitem{ref44}
Q. Wu, Y. Zeng, and R. Zhang, ``Joint trajectory and communication design for multi-UAV enabled wireless networks,'' \emph{IEEE Trans. Wireless Commun.}, vol. 17, no. 3, pp. 2109–2121, Mar. 2018.

\bibitem{refmanopt}
N. Boumal, An introduction to optimization on smooth sanifolds, 1st ed. Cambridge, U.K.: Cambridge Univ. Press, 2023.

\end{thebibliography}
\end{document}